\documentclass[aps,pra,reprint,twocolumn]{revtex4-2}
\usepackage[T1]{fontenc} 
\usepackage[utf8]{inputenc} 
\usepackage{amsmath,amssymb,bm}
\usepackage{graphicx}
\usepackage{xcolor} 
\usepackage{amsthm}
\usepackage{hyperref}

\begin{document}
\newtheorem{theorem}{Theorem}
\newtheorem{corollary}{Corollary}
\newtheorem{proposition}{Proposition}
\newtheorem*{customprop}{Proposition} 
   
\def\be{\begin{equation}}
\def\en#1{\label{#1}\end{equation}}
 \def\S{\mathcal{S}}
\def\D{\mathcal{D}}
 \def\vare{\varepsilon}
\newcommand{\per}{\mathrm{per}}

\title{  Indistinguishability  Theory  for  Identical Particles with  Invariant Degrees of Freedom  }

\author{Valery Shchesnovich }
\affiliation{Centro de Ci\^encias Naturais e Humanas, Universidade Federal do
ABC, Santo Andr\'e,  SP, 09210-170 Brazil }
\date{\today}

\begin{abstract}
Identical  particles  have  dynamically invariant labels  in all  setups where   their  degrees of freedom  can be  partitioned  into two parts, which we call internal and visible, with the visible part  subject to unitary evolution, while the dynamically invariant internal part  accounts for the partial distinguishability of particles. We give the explicit form of the visible state in terms of the indistinguishability function, generalizing the  standard symmetrization/antisymmetrization postulate for bosons/fermions with internal modes.      The  indistinguishability function  on the symmetric group accounts for the permutation symmetry of the dynamically invariant label state and gives a   self-contained description of  the  symmetry spectrum of the visible state.  It is shown that the indistinguishability of single particles emitted independently from a stable source   is completely characterized  by  the   projective  measures on  the   symmetry spectrum.  We   reveal a  hierarchy of successive upper bounds on the projective measure of indistinguishability in terms of marginal projective measures. We  point out an experimentally feasible way to get the upper bounds on the indistinguishability of an arbitrarily large number of single photons  by  direct experimental readout of their  marginal projective measures.  We also analyze the indistinguishability of coherent superpositions and convex mixtures and resolve a recently posed problem. \end{abstract}

\maketitle

\section{Introduction}
\label{sec1}
The concept of partial indistinguishability originates from the Hong–Ou–Mandel two-photon interference experiment \cite{HOM}, where the visibility of the coincidence counts is determined by the degree of exchange symmetry of the two-photon state. Permutation symmetry of the internal state likewise governs interference phenomena involving two fermions \cite{ElecHOM}, two atoms \cite{AtomHOM}, and quantum walks of two entangled photons \cite{TwFerQW}. More generally, permutation symmetries of multiphoton states determine their interference and bunching properties \cite{Ou1,Ou2}. Multiparticle interference of identical bosons and fermions exhibits both statistics-dependent effects \cite{GenHOM} and phenomena arising from more general symmetry principles \cite{SymBeyBS,ZeroTranSymm,Symm4Dist,MetHOM}.     Indistinguishability  of identical particles also unlocks   entanglement-based  quantum information  protocols  \cite{LF1,LF2,LF3}. In general, entanglement in the context of  systems of identical particles  remains a complex issue  \cite{RepEnt}.   

A complete characterization of partial distinguishability for more than two identical particles remains a surprisingly challenging problem \cite{Shch2015,Tch2015,WeylD}. Genuinely multiparticle effects are remarkably abundant and include non-monotonic quantum-to-classical transitions \cite{NonMon4ph}, collective multiparticle phases \cite{3phPhase,DistMix3ph,nphPhases}, interference involving subsets of distinguishable particles \cite{DistPhInter}, efficient methods for characterizing multiparticle indistinguishability \cite{MultPhInd,Distchar}, and  non-trivial     bunching  properties \cite{VS2016,BCount,Geller2026}.

These effects preclude using a single scalar measure to order all interference phenomena universally. The most important measure of indistinguishability is the probability of the ideal case, given by the projection onto the symmetric subspace of the Hilbert space of internal states \cite{Shch2014,Shch2015}, also considered recently within the quantum information framework \cite{DistNew}. This quantity  is the complement of an upper bound on the distinguishability error \cite{Shch2015A} in Boson Sampling \cite{AA}, where large-scale multiparticle interference \cite{20ph60mod} must compete with increasingly powerful classical simulation methods \cite{SimBSdist}. It is expected to play a similarly important role in photonic approaches to universal quantum computation \cite{LOC,RevLOC}, which rely on the indistinguishability of photons. 

The purpose of this  work is to provide a framework for   partial indistinguishability of identical particles having dynamically invariant degrees of freedom. We restrict our approach to unitary linear evolutions,   realized, for example, by spatial unitary interferometers for identical particles. The  key  ingredient in our approach is the ``label state'' of identical particles, which is a dynamical invariant. The      indistinguishability function  encodes the information on the  label state and   defines the visible state of identical particles, i.e., the state operated on by the evolution. We give the explicit expression for the visible state of identical particles in orthogonal visible modes  in terms of the indistinguishability function and  analyze     coherent superpositions of such states.    We relate  the projective indistinguishability measure and the emulation measure of one species by the other to the  dynamically invariant  label state. Furthermore,  we  characterize class-function indistinguishability and, for specified orthogonal visible modes and a given occupation vector, the corresponding visible state, by the projective measures on the generalized symmetry sectors. We introduce marginal projective measures of indistinguishability, which are an ordered hierarchy of upper bounds on the projective measure of indistinguishability. We  reveal how  the upper bounds can be read out in an    interference experiment  with   bosons.  Finally, we present nontrivial partially distinguishable states sharing the projective indistinguishability measure of maximally distinguishable particles, thus resolving a recently posed problem. 

The paper is organized as follows. In Section~\ref{sec2}, Theorem~1 and Corollary~1 connect the projective measures of indistinguishability and of emulation of one particle species by the other, respectively, to the internal state of identical particles. In Section~\ref{sec3}, Theorem~2 and Proposition~1 express the visible state of identical particles in terms of the indistinguishability function. We also discuss the effect of a linear interferometer, maximally distinguishable particles, and the indistinguishability of coherent superpositions of states. In Section~\ref{sec4}, the indistinguishability function is analyzed using generalized permutation symmetries. Proposition~2 gives an expansion for indistinguishability functions that are also class functions on the symmetric group, while Theorem~3, which generalizes Theorem~1 and Corollary~1, connects the expansion coefficients to projective measures of the generalized symmetry sectors. We characterize maximal distinguishability by the Plancherel distribution and, for non-class indistinguishability functions, connect the indistinguishability-function description to a matrix-valued symmetry spectrum. In Section~\ref{sec5}, we introduce marginal states and marginal projective measures. Theorem~4 shows that these measures form a hierarchy of upper bounds on the projective indistinguishability measure, while Proposition~3 supplies a lower bound for convex mixtures of tensor-power label states. We also propose a method for obtaining the marginal projective measures of identical bosons directly from an interference experiment. Section~\ref{sec6} summarizes the results. Mathematical details and most derivations are collected in Appendices~\ref{appA}--\ref{appD}.
\section{Projective measures   of symmetry  }
\label{sec2}

We consider $n$ identical particles, bosons or fermions,  and partition the single-particle Hilbert space as $\mathcal{H}=\mathcal{H}_{(vis)}\otimes \mathcal{H}_{(int)}$, where the visible (operated-on) degrees of freedom are subject to the same single-particle unitary on every particle (i.e., to a unitary linear  interferometer) while the internal degrees of freedom remain invariant. Thus the complete evolution is $(\hat U\otimes\hat I_{(int)})^{\otimes n}$ (see also Section \ref{sec3B} below). Accordingly,  we assume a finite $\dim\mathcal H_{(vis)}=M$.  
We will mostly present our results in  the first-quantization representation (as we   trace out part of the degrees of freedom of the particles), resorting to the second-quantization representation when necessary  (for the precise mathematical relation between the two representations, see, e.g., Ref. \cite{LecNotes}). Recall that in the first-quantization representation the state of $n$ identical bosons/fermions is symmetric/antisymmetric. The projectors onto the symmetric/antisymmetric subspaces of   $\mathcal{H}^{\otimes n}$ read:
\be
\hat{S}^{(\pm)} := \frac{1}{n!}\sum_{\sigma\in S_n} \vare(\sigma) \hat{P}_\sigma, \quad \vare(\sigma) = \left\{ \begin{array}{cc} 1, & \mathrm{bosons},\\ \mathrm{sgn}(\sigma), & \mathrm{fermions},\end{array} \right.
\en{Svare}
where the sum runs over the  symmetric group $S_n$ of $n$ objects, $\mathrm{sgn}(\sigma)$ is the signature of the permutation $\sigma$  and 
\be
\hat{P}_\sigma\bigotimes_{j=1}^n |\phi_j\rangle:= \bigotimes_{j=1}^n |\phi_{\sigma^{-1}(j)}\rangle
\en{Pdef}
is  the  unitary representation of $\sigma$ in    $\mathcal{H}^{\otimes n}$.  Below we will also use the corresponding permutation operators in the visible/internal subspaces and use the same notation, where it does not lead to confusion, bearing in mind that  the operator in Eq.~(\ref{Pdef}) is a tensor product of the latter $\hat{P}_\sigma = \hat{P}_\sigma\otimes \hat{P}_\sigma$. 

The unitary linear evolution (a unitary linear interferometer) affects only the visible part, leaving the internal part invariant. These are defined as follows: 
\be
\hat{\varrho}_{(vis)}=\mathrm{Tr}_{(int)} \hat{\varrho},  \quad \hat{\varrho}_{(int)}=\mathrm{Tr}_{(vis)}\hat{\varrho}.
\en{visST_def}
Both parts   possess    the particle permutation symmetry    
\be
 \hat{P}_\sigma\hat{\varrho}_{(vis)}\hat{P}^\dag_\sigma=\hat{\varrho}_{(vis)}, \quad \hat{P}_\sigma\hat{\varrho}_{(int)}\hat{P}^\dag_\sigma=\hat{\varrho}_{(int)}, \quad \forall \sigma\in S_n,
 \en{permvisST}
   which follows from the same symmetry of the complete state $\hat{\varrho}$. 
      
 We are interested in the probability   $p_{\bm{m}}$ of observing an occupation vector  $\bm{m}=(m_1,m_2,\ldots,m_M)$, $|\bm{m}|:=m_1+m_2+\ldots =n$, in some  basis of   visible modes.   Irrespective of whether the internal modes are resolved  at the particle counting detection stage,    the probability  $p_{\bm{m}}$ is completely determined by   $\hat{\varrho}_{(vis)}$.  Thus, we can use the  particle-number-counting detection operator   of  occupation vector $\bm{m}$ without  resolution of  the internal modes (with the internal data being  traced out). The latter   is easy to  derive from first principles \cite{Shch2014,Shch2015}.   The resulting detection operators are   
 \be
  \hat{\Pi}^{(\pm)}_{\bm{m}}:=\hat{S}^{(\pm)}\left(\hat{\Pi}_{\bm{m}} \otimes \hat{I} \right)\hat{S}^{(\pm)} 
  \en{fulDetOp}
  with 
\be
\hat{\Pi}_{\bm{m}}\equiv \frac{1}{\bm{m}!}\sum_{\sigma\in S_n}\bigotimes_{\alpha=1}^n|\tilde{\ell}_{\sigma(\alpha)}\rangle\langle \tilde{\ell}_{\sigma(\alpha)}|,\quad \sum_{\bm{m}}\hat{\Pi}_{\bm{m}}=\hat{I},
\en{DetOp}
  where the states $|\tilde{\ell}\rangle$, $\ell=1,\ldots,M$,   are  the (particle counting) detection  basis of $\mathcal{H}_{(vis)}$, and the denominator 
  $  \bm{m}!:= m_1!\ldots m_M!$     accounts  for multiple counting of the same output modes by  the  permuted sequences.  
  
Since the detection operators in Eq.~(\ref{fulDetOp}) apply to complete   states  already  properly symmetrized/antisymmetrized, we can  use instead of the   detection operators    of Eq.~(\ref{fulDetOp}) the simpler product operators $ \hat{\Pi}_{\bm{m}} \otimes \hat{I}$,  or    the   operators in Eq. (\ref{DetOp}) applied directly  to the visible state Eq.~(\ref{visST_def}).    Therefore, the  probability of the occupation vector $\bm{m}$  becomes   
\be
  p_{\bm{m}}=\mathrm{Tr}\{ \hat{\Pi}_{\bm{m}} \hat{\varrho}_{(vis)}\}.
\en{pm}
 It is important   that  an appropriate choice of  the particle counting detection basis $|\tilde{\ell}\rangle\in \mathcal{H}_{(vis)}$ in the detection operator $\hat{\Pi}_{\bm{m}}$ in Eq.~(\ref{DetOp}) can also account  for the action of an arbitrary linear unitary interferometer   on the state $\hat{\varrho}_{(vis)}$ when computing the probability by Eq.~(\ref{pm}) (see also Section \ref{sec3B} below).

Indistinguishability  of identical particles with invariant internal degrees of freedom can be estimated by  comparing  $p_{\bm{m}}$ with the corresponding ideal case, i.e., the corresponding distribution  $p^{(i)}_{\bm{m}}$  which would be obtained with completely  indistinguishable particles (for example, identical particles  with  a one-dimensional internal subspace $\mathcal{H}_{(int)}$).    As the  ideal counterpart of  $\hat{\varrho}$ we can take the state   $\hat{\varrho}^{(i)}\equiv \hat{\varrho}_{(vis)}^{(i)}\otimes \left(|\psi\rangle\langle\psi|\right)^{\otimes n} $, with an arbitrary $|\psi\rangle$ and its visible part given by
\be
\hat{\varrho}_{(vis)}^{(i)} :=\frac{\hat{S}^{(\pm)} \hat{\varrho}_{(vis)} \hat{S}^{(\pm)}}{\mathrm{Tr}\left( \hat{S}^{(\pm)} \hat{\varrho}_{(vis)} \right)}.
\en{varrho_i}
 Eq.(\ref{varrho_i})  presupposes  that $\mathrm{Tr}(\hat S^{(\pm)}\hat\varrho_{(vis)})>0$.
This assumption  fails for bosons in an  antisymmetric visible state,  and, respectively, for     fermions in  a symmetric visible state. In such a case one species emulates the other (see below).  
  
The maximal total variation distance between the probability distributions $p_{\bm{m}}$ and $p^{(i)}_{\bm{m}}$ corresponding to the ideal counterpart  is known \cite{BookNC} to be bounded by   the trace distance between the visible states:
\begin{eqnarray}
 \underset{\hat{\Pi}_{\bm{m}}} {\mathrm{max} }
\left\{\frac12\sum_{\bm{m}}| p_{\bm{m}} - p^{(i)}_{\bm{m}}|\right\}    \le  \frac12\mathrm{Tr}\left|\hat{\varrho}_{(vis)}-\hat{\varrho}_{(vis)}^{(i)}\right| . 
\label{Trd}
\end{eqnarray}
Here the maximization is restricted to particle-counting measurements after a linear unitary  interferometer  (hence, the upper bound in general).  

There is another  physically transparent expression for  the trace distance. To this end, we now introduce the indistinguishability measure $\mathcal{D}(\hat{\varrho})$ as the complement of the trace distance in  Eq.~(\ref{Trd}):
\be 
 \mathcal{D} (\hat{\varrho}):=1-  \frac12\mathrm{Tr}\left|\hat{\varrho}_{(vis)}-\hat{\varrho}_{(vis)}^{(i)}\right|  .
 \en{Trd_ind}
We have the following result (proven  in Appendix~\ref{appA}). 
\begin{theorem}
\be
\mathcal{D} (\hat{\varrho}) = \mathrm{Tr}\left( \hat{S}^{(\pm)}\hat{\varrho}_{(vis)} \right) =\mathrm{Tr}\left( \hat{S}^{(+)}\hat{\varrho}_{(int)}\right).
\en{Dint}
The projective  measure  $\mathcal D$ is defined for every state, while the trace-distance expression in Eq.~(\ref{Trd_ind}) applies when Eq.~(\ref{varrho_i}) is defined. 
\end{theorem}
Equation~(\ref{Dint}) gives the probability that the visible state of bosons or fermions belongs to the ideal symmetric/antisymmetric subspace (whether or not the normalized ideal counterpart exists). For bosons, the first  expression in Eq.~(\ref{Dint}) was adopted in Ref.~\cite{DistNew}, whereas the second was  utilized  in Refs.~\cite{Shch2015,Shch2015A}.

By appropriately entangling the internal degrees of freedom, bosons can emulate fermionic behavior and vice versa \cite{TwFerQW,BSF} (e.g.,  the zero-denominator case of Eq.~(\ref{varrho_i})).  The proof of Theorem~1  also points   to an   analogous measure for such an emulation (see Appendix~\ref{appA}).
\begin{corollary}
The following projective measure       
\be
\widetilde{\mathcal{D}}(\hat{\varrho}) := \mathrm{Tr}\left( \hat{S}^{(\mp)}\hat{\varrho}_{(vis)} \right) =\mathrm{Tr}\left( \hat{S}^{(-)}\hat{\varrho}_{(int)}\right)
\en{emD}
gives the projection probability onto the opposite exchange-symmetry sector and therefore quantifies the degree of  emulation of one species by the other.
\end{corollary}
 The  physical significance and generalizations of the two dynamically invariant measures in Eqs.~(\ref{Dint})-(\ref{emD})  are  explored below by using the indistinguishability theory.
 
\section{Visible state, label state, and    indistinguishability function}
\label{sec3}

The second expression in Eq.~(\ref{Dint})  first appeared within the   partial indistinguishability  theory  in Refs.~\cite{Shch2014,Shch2015,Shch2015A,VS2016}, where the key  object   is the indistinguishability function of identical particles
\be
J_{\hat{\varrho}}(\sigma)
:=
\mathrm{Tr}(\hat{P}_\sigma\hat{\varrho}^{(l)}),
\en{distJ}
where $\hat{\varrho}^{(l)}$ is the state of particle labels  associated with $\hat{\varrho}$ (we have changed the  nomenclature here:   in Refs.~\cite{Shch2014,Shch2015,Shch2015A,VS2016} the state $\hat{\varrho}^{(l)}$  was called the internal state;  here the term ``internal state'' is already reserved for the reduced state introduced in Eq.~(\ref{visST_def})).  

In general, many different label states correspond to the same indistinguishability function. In order to   reconstruct  the   label state  one would need internal-state-resolving particle counting detection.  Whether such a   reconstruction is   feasible or not, the indistinguishability function is all that is required for particle counting detection (or when information on  the internal state is  traced out; see  also Ref.  \cite{nphPhases}), as it contains all information about $\hat{\varrho}$ relevant for the description of interference in unitary linear interferometers, for fixed orthogonal input modes and a fixed occupation vector \cite{Shch2014,Shch2015}. For one particle per occupied input mode, the full indistinguishability function can be reconstructed with a suitable family of interferometers, including auxiliary vacuum modes when needed \cite{PartDistInv}. Cyclic interferometers give access to particular cycle contributions \cite{MultPhInd}.  

Below we reveal how the indistinguishability function encodes the visible state for fixed orthogonal input modes and a fixed occupation vector.   
 \subsection{Visible state and indistinguishability function}
 \label{sec3A}

 Now we introduce the label state  and derive   the visible part of  the  pure   state $\hat{\varrho} =|\Psi_{\bm{n}}\rangle \langle \Psi_{\bm{n}}|$ of $n$ identical  particles occupying  $r\le \mathrm{min}(n,M)$ orthonormal visible modes        $|k\rangle \in \mathcal{H}_{(vis)}$, $k=1,\ldots,M$, with $n_k$ particles in mode $k$ (here $n_k\ge 0$). Then 
\be	
 |\Psi_{\bm{n}}\rangle = \sum_{j_1,\ldots,j_n} C_{j_1,\ldots,j_n} \frac{\prod_{\alpha=1}^n \hat{a}^\dag_{k_\alpha,j_\alpha}}{\sqrt{\bm{n}!}}|0\rangle,
\en{Psi_nSQ}
where   $k_1\le \ldots\le  k_n$ is the  sequence  of modes corresponding to the occupation vector $\bm{n}\equiv (n_1,\ldots,n_M)$ (in the case of fermions, there is a preset order in the product),  $\bm{n}! = n_1!\ldots n_M!$,   $|j\rangle$, $j=1,2,3,\ldots$, is a  basis in $\mathcal{H}_{(int)}$, and the boson/fermion operator $\hat{a}^{\dag}_{k,j}$ creates a particle in the state $|k,j\rangle$. Due to the permutation symmetry/antisymmetry of the creation operators, the coefficients   $C_{j_1,\ldots,j_n}$  can be chosen symmetric/antisymmetric with respect to the Young subgroup 
 \be 
 \mathcal{Y}_{\bm{n}}\equiv S_{n_1}\times S_{n_2}\times \ldots \times S_{n_M}
 \en{YoungSG}
(if $n_k=0$, $S_{n_k}$ is omitted), namely 
\be
C_{j_{\sigma(1)},\ldots,j_{\sigma(n)}}  = \vare(\sigma) C_{j_1,\ldots,j_n}, \quad \forall \sigma \in \mathcal{Y}_{\bm{n}}.
\en{ED8}
Under the symmetry condition in Eq.~(\ref{ED8}), the normalization condition  reads  
\[
\sum_{j_1,\ldots,j_n}|C_{j_1,\ldots,j_n}|^2 = 1.  
\]
Partial indistinguishability can be  encoded by the permutation properties of a label state in  the internal subspace $\mathcal{H}_{(int)}^{\otimes n}$. For the $n$-particle state of Eq. (\ref{Psi_nSQ})   the label state is defined as follows    
\be
\hat{\varrho}^{(l)} := |\Psi^{(l)}_{\bm{n}}\rangle \langle \Psi^{(l)}_{\bm{n}}|,\quad |\Psi^{(l)}_{\bm{n}}\rangle:= \sum_{j_1,\ldots,j_n}C_{j_1,\ldots,j_n}\bigotimes_{\alpha=1}^n |j_\alpha\rangle.
\en{Psi_nlab}
It is important to note that, in general,    $\hat{\varrho}^{(l)}\ne \hat{\varrho}_{(int)}$. Indeed,  Eqs.~(\ref{visST_def}) and (\ref{Psi_nSQ}) give 
\be
  \hat{\varrho}_{(int)} = \mathrm{Tr}_{(vis)}(|\Psi_{\bm{n}}\rangle \langle \Psi_{\bm{n}}|) = \frac{1}{n!}\sum_{\sigma \in S_n} \hat{P}_\sigma  \hat{\varrho}^{(l)} \hat{P}^\dag_\sigma.
  \en{symrhoint} 
Thus the internal state can be understood as the label state where the  unique association between the labels and the particles is erased by  permuting the individual (indexed by $\alpha$ in Eq. (\ref{Psi_nlab})) single-particle Hilbert spaces $\mathcal{H}_{(int)}$ in the tensor power     $\mathcal{H}_{(int)}^{\otimes n}$. It is important to note that  while the   internal state  is always invariant under the particle permutations Eq.~(\ref{permvisST}),   the label state, in general, has no invariances  under  particle permutations. One mandatory symmetry  of the label state occurs for  multiple occupations of the visible modes: in this case it inherits the Young-subgroup symmetry   Eq.~(\ref{ED8}).

Consider now the  indistinguishability function, Eq.~(\ref{distJ}), for the state of Eq. (\ref{Psi_nSQ}). It  inherits the Young-subgroup symmetry of Eq.~(\ref{ED8}); we have 
\be
  J_{\hat{\varrho}}(\sigma\pi) =  J_{\hat{\varrho}}(\pi\sigma)=\vare(\pi) J_{\hat{\varrho}}(\sigma), \quad \forall \pi \in \mathcal{Y}_{\bm{n}}. 
\en{JsymY}
We have  introduced  all the prerequisites necessary to give the corresponding   visible state (see the proof in Appendix~\ref{appB}).
\begin{theorem}
The visible component of the state in  Eq.~(\ref{Psi_nSQ}) is
\begin{eqnarray}
\label{vis_state}
&& \hat{\varrho}_{(vis)} :=\mathrm{Tr}_{(int)}\left(|\Psi_{\bm{n}}\rangle \langle \Psi_{\bm{n}}|\right)\nonumber\\
&& = \frac{1}{n!\bm{n}!} \sum_{\sigma,\pi\in S_n} \vare(\pi\sigma)  J_{\hat{\varrho}}(\pi\sigma^{-1})  \bigotimes\limits_{\alpha=1}^n | k_{\sigma(\alpha)}\rangle\langle k_{\pi(\alpha)}|,\nonumber\\
 \end{eqnarray} 
 where $\hat{\varrho}^{(l)}$ is   defined in Eq.~(\ref{Psi_nlab}).
Conversely,  every  visible state of bosons/fermions  with the   occupation vector   $\bm{n}=(n_1,\ldots, n_M)$, with the  corresponding   sequence of modes  $k_1\le \ldots\le  k_n$,   has the form of Eq.~(\ref{vis_state}) for  some   positive semidefinite  function $J(\sigma)$,  
  \be
  \sum_{\sigma,\pi\in S_n} Z^*_\pi J(\pi\sigma^{-1}) Z_\sigma \ge 0, \quad \forall Z_\sigma\in \mathbb{C}, 
  \en{psdJ}
   satisfying  Eq.~(\ref{JsymY}) and $J(e)=1$ ($e$  is  the  identity in $S_n$).    
\end{theorem}
One can verify that the   state of Eq.~(\ref{vis_state})    satisfies the   particle permutation symmetry of the visible state Eq.~(\ref{permvisST}),
  where the action of the particle permutation  amounts to reordering of the two permutations $\sigma\to  \sigma\tau$ and $\pi \to  \pi\tau$   not affecting the  indistinguishability function (recall that $\vare(\sigma^{-1}) = \vare(\sigma)$, the function being either constant or the  signature of a  permutation).  

The  indistinguishability  function  defined in   Eq.~(\ref{distJ}) is  a properly normalized   positive semidefinite  function on the symmetric group $S_n$, i.e., it satisfies Eq. (\ref{psdJ})  by definition  and by  the group property $\hat{P}_{\sigma\pi} = \hat{P}_\sigma \hat{P}_\pi$.  The  converse  result holds (see Appendix \ref{appB}).
\begin{proposition}
Every  positive semidefinite  function $J(\sigma)$ on $S_n$, satisfying also $J(e)=1$, can be cast in the  form of the indistinguishability function, i.e., 
\be
J(\sigma) = \mathrm{Tr}(\hat{P}_\sigma \hat{\varrho}^{(l)})
\en{Jform} 
for some  label  state $\hat{\varrho}^{(l)}\in \mathcal{H}_{(int)}^{\otimes n}$ when  $\mathrm{dim}(\mathcal{H}_{(int)})\ge n$.  
\end{proposition}
If the  dimension of the  single-particle internal Hilbert space is less than $n$, the physical indistinguishability function is still given by   Eq.~(\ref{Jform}), but not all possible positive semidefinite functions on $S_n$ can be realized  in this form (for instance, the trivial indistinguishability function  $J(\sigma) = \delta_{\sigma,e}$, where $e$ is the identity in $S_n$,   requires $\dim\mathcal H_{(int)}\ge n$).

 Observe that the visible state  $\hat{\varrho}_{(vis)}$  in Theorem 2  gets a   uniform  expression valid for the two species, bosons and fermions, upon the  introduction of the uniform  positive semidefinite function   $\Lambda(\sigma) =  \vare(\sigma)J(\sigma)$.

By using  the  identity 
 \be
\hat{P}_\sigma\hat{S}^{(\pm)} =\hat{S}^{(\pm)}   \hat{P}_\sigma= \vare(\sigma) \hat{S}^{(\pm)}, \quad \forall \sigma\in S_n,
 \en{PS}
  one can substitute the label state for the internal state in the projective measures of Theorem 1 and Corollary 1 of Section \ref{sec2}. Then, if we happen to know  the physical setup of  preparation of  $n$ identical particles,  we can compute the respective projective measures from the label state \cite{Shch2014,Shch2015,Shch2015A}.  
From Eqs.~(\ref{Dint}), (\ref{emD}) and (\ref{vis_state}), using  $\langle l|k\rangle=\delta_{l,k}$ and   $ \mathrm{sgn}(\sigma^{-1}) = \mathrm{sgn}(\sigma)$,  one readily obtains
\be
  \mathcal{D}   = \frac{1}{n!}\sum_{\sigma\in S_n} J_{\hat{\varrho}}(\sigma), \quad  \widetilde{\mathcal{D}}  = \frac{1}{n!}\sum_{\sigma\in S_n}  \mathrm{sgn}(\sigma)J_{\hat{\varrho}}(\sigma).
\en{DJ}

As an example, consider the visible state of $n$ identical bosons/fermions occupying the same visible mode $|1\rangle$, i.e.,  when their   label state is   symmetric in the case of bosons and   antisymmetric in the  case of fermions. Indeed, in this case  $\mathcal{Y}_{\bm{n}}= S_n$ and Eq.~(\ref{JsymY}) gives $J_{\hat{\varrho}}(\sigma) = \vare(\sigma)$. Hence,  by   Eq.~(\ref{DJ})  bosons are  completely indistinguishable in this case ($\mathcal{D}=1$), while fermions ideally emulate bosons ($\widetilde{\mathcal{D}}=1$).   From  Eq.~(\ref{vis_state}) we obtain 
\be
 \hat{\varrho}_{(vis)} =     \left(| 1\rangle\langle 1|\right)^{\otimes n}. 
 \en{one-mode}
A  similar  state can apply to  \textit{non-identical} particles, so an interference experiment with identical particles  in the state of Eq.~(\ref{one-mode}) would resemble that of non-identical particles.  However,   this does not imply the   maximal  distinguishability of identical particles,  as, e.g.,  bosons in the state of Eq.~(\ref{one-mode})  are completely indistinguishable due to their   completely symmetric label state.  The precise criterion for the maximal distinguishability is introduced below (see Section \ref{sec3C}).

 \subsection{The effect of  a linear  interferometer}
 \label{sec3B}
 
Now  we can analyze the effect of a linear unitary interferometer $\hat{U}$ which    acts on the visible modes $k=1,\ldots,M$. In the second-quantization representation  its action is  as follows 
\be
\hat{a}^\dag_{k,j} = \sum_{\ell=1}^MU_{k,\ell} \hat{b}^\dag_{\ell,j},
\en{Uab2}
where $\hat{b}^\dag_{\ell,j}$ is the  creation operator of a boson/fermion in an output mode $\ell$ and   internal mode $j$,  i.e., in a state 
$|\tilde{\ell},j\rangle=|\tilde{\ell}\rangle\otimes|j\rangle$, $|\tilde{\ell}\rangle\in \mathcal{H}_{(vis)}$ (here  $|j\rangle\in \mathcal{H}_{(int)}$).  
Accordingly, in the first-quantization representation the action of the interferometer of Eq.~(\ref{Uab2}) is represented by the expansion of the visible basis states in the output-mode basis 
\be
|k\rangle = \sum_{\ell=1}^M U_{k,\ell} |\tilde{\ell}\rangle. 
\en{Uab1}
Applying Eq.~(\ref{Uab1}) to the state of Eq.~(\ref{vis_state})  we  obtain    the  visible diagonal matrix elements  
\be
\varrho_{\bm{\ell},\bm{\ell}}:=   \langle \tilde{\ell}_1|\otimes \ldots \otimes \langle \tilde{\ell}_n| \hat{\varrho}_{(vis)} 
|\tilde{\ell}_1\rangle\otimes \ldots \otimes |\tilde{\ell}_n\rangle
\en{rhoell}
accessible by  particle counting detection  (see Eq.~(\ref{DetOp})) in the following form  
\begin{eqnarray}
 \!\! \varrho_{\bm{\ell},\bm{\ell}} = \frac{1}{n!\bm{n}!}  \!\!\sum_{\sigma,\pi\in S_n}  \!\! \vare(\pi\sigma)    J_{\hat{\varrho}}(\pi\sigma^{-1})  \!\! \prod_{\alpha=1}^n U_{k_{\sigma(\alpha)},\ell_\alpha}U^*_{k_{\pi(\alpha)},\ell_\alpha}.\nonumber\\
 \label{rho_ell}
 \end{eqnarray}
  The expression in Eq.~(\ref{rhoell})  for the matrix element is obviously invariant under  the particle permutation symmetry Eq.~(\ref{permvisST}) of the visible state. This can also be verified  directly in  Eq.~(\ref{rho_ell}), where   the simultaneous permutation $\tau$ of the output modes $\bm{\ell}$ in $U$ and $U^*$    can be absorbed by  reordering   the two permutations   $\sigma\to\sigma\tau $ and  $\pi\to \pi\tau $ (which  leaves  the function $\vare(\pi\sigma) J(\pi\sigma^{-1})$ invariant),   with simultaneous  reordering of the  scalar product.

 Taking the trace with  the detection operator of Eq.~(\ref{DetOp}) amounts to multiplying the diagonal matrix element $\varrho_{\bm{\ell},\bm{\ell}}$ Eq.~(\ref{rhoell}) by the multinomial factor ${n!}/{\bm{m}!}$ (where $n!$ is due to simultaneous permutations of the output modes $\ell_\alpha$, as discussed above). Hence, the  output probability $p_{\bm{m},\bm{n}}$ of detecting identical particles in an  output occupation vector $\bm{m}$ with the  input vector $\bm{n}$ becomes  \cite{Shch2015,VS2016} 
\be
p_{\bm{m},\bm{n}} = \frac{1}{\bm{m}!\bm{n}!} \!\!\sum_{\sigma,\pi\in S_n}  \!\! \vare(\pi\sigma)    J_{\hat{\varrho}}(\pi\sigma^{-1})  \!\! \prod_{\alpha=1}^n U_{k_{\sigma(\alpha)},\ell_\alpha}U^*_{k_{\pi(\alpha)},\ell_\alpha}.
\en{prob_{nm}}

The output probability  is a function of  two permutations, in general. One permutation  accounts for  identical particles, whatever their state of indistinguishability (originating from the particle permutation  symmetry of the  visible state  Eq.~(\ref{permvisST})), while   the relative permutation, the argument of the    indistinguishability function, accounts for the  multi-particle interference in   the  output  quantum amplitude.
Accordingly, the  matrix elements over the output modes in Eq.~(\ref{rho_ell}) depend on both the input indistinguishability function and the interferometer. 

For  the particle counting  detection acting  as a post-selection, we can introduce,  for  any  $p_{\bm m,\bm n}>0$,   the  post-selected    state  over the  occupation sector  $\bm m$ as follows 
\be
 \hat\varrho_{(vis)|\bm m}
 =\frac{\hat\Pi_{\bm m}\hat\varrho_{(vis)}\hat\Pi_{\bm m}} {p_{\bm m,\bm n}}.
\en{visST_red}
This  post-selected state  is  invariant under the particle permutations Eq.~(\ref{permvisST}). By the converse part of Theorem~2, it therefore admits the form of Eq.~(\ref{vis_state}) with  a post-selected indistinguishability function  $J_{\bm{m}}$  (which depends on the post-selected  sector $\bm{m}$ through  the corresponding  matrix elements $U_{k,l}$).   By direct computations using Eqs.~(\ref{vis_state}) and (\ref{visST_red}) we get the   post-selected indistinguishability function as follows 
 \begin{eqnarray}
&& J_{\bm{m}}(\tau) = \vare(\tau) \frac{\mathcal{Z}_{\bm{m}}(\tau)}{\mathcal{Z}_{\bm{m}}(e)},\\
&& \mathcal{Z}_{\bm{m}}(\tau):= \!\!\sum_{\sigma,\pi\in S_n}  \!\! \vare(\pi\sigma)    J_{\hat{\varrho}}(\pi\sigma^{-1})  \!\! \prod_{\alpha=1}^n U_{k_{\sigma(\alpha)},\ell_\alpha}U^*_{k_{\pi(\alpha)},\ell_{\tau(\alpha)}}.\nonumber
\end{eqnarray}
In general,  $J_{\bm{m}}$ differs from the input indistinguishability function $J_{\hat{\varrho}}$. For instance, $J_{\bm{m}}$ inherits the  permutation symmetry with respect to the  Young subgroup of the post-selected output sector \mbox{$\mathcal{Y}_{\bm{m}} = S_{m_1}\times \ldots \times S_{m_M}$}:
\be
  J_{\bm{m}}(\tau\pi) =  J_{\bm{m}}(\pi\tau)=\vare(\pi) J_{\bm{m}}(\tau), \quad \forall \pi \in \mathcal{Y}_{\bm{m}}. 
\en{Jm_symY}

\subsection{Maximally distinguishable identical particles}
\label{sec3C}

Our notion of maximal distinguishability of identical particles generalizes the vanishing-overlap condition of the Hong--Ou--Mandel effect \cite{HOM}. It corresponds to mutually orthogonal particle labels in the simplest case, i.e., the label state is a product of mutually orthogonal states. More generally it corresponds to  the trivial indistinguishability function \cite{Shch2015}
\be J^{(d)}(\sigma) :=\delta_{\sigma,e}
\en{Jmaxdist}
(here $e$ is the identity in $S_n$).
 The physical basis of this criterion is the absence of exchange interference; thus, it generalizes the distinguishable-particle limit of the Hong--Ou--Mandel experiment.
For two bosons incident in distinct input ports of a balanced beam splitter  the coincidence probability is
\be
 p_{(1,1),(1,1)}=\frac{1-J((12))}{2}.
\en{HOMcriterion}
For independent pure labels, $J((12))=|\langle\phi_1|\phi_2\rangle|^2$, so orthogonal labels give $p_{(1,1),(1,1)}=1/2$ and remove the exchange contribution. For $n$ particles in distinct orthogonal input modes, Eq.~(\ref{Jmaxdist}) removes every term with $\pi\ne\sigma$ from Eq.~(\ref{prob_{nm}}), leaving the incoherent sum of particle-assignment probabilities.   The condition in Eq.~(\ref{Jmaxdist}) concerns the particle exchange interference on a single interferometer and not the multi-particle  interference  when the visible  modes are superpositions of the interferometer input modes  (the coherent superpositions of visible states   are  analyzed in  Section \ref{sec3D} below).

Our criterion in Eq.~(\ref{Jmaxdist}), in a general setting,   differs from the  maximal  distinguishability criterion for bosons of Ref.~\cite{DistNew}.  The distinction originates from   the defining physical criteria for maximal distinguishability and, therefore,  leads to  different properties under mixing  (see also the discussion in Section \ref{sec3D} below). 

For Eq.~(\ref{Jmaxdist}), we have from Eq.~(\ref{DJ})
\be
\widetilde{\mathcal{D}}=\mathcal{D}=\frac{1}{n!}.
\en{Dmaxdist}
However, Eq.~(\ref{Dmaxdist})  applies to a far more general set of states than just the maximally distinguishable states  (we  analyze this   in  Section \ref{sec4} below).

For an $n$-particle state $\hat{\varrho}$ with the occupation vector $\bm n$ containing multiply occupied modes, the indistinguishability function has the Young-subgroup symmetry Eq.~(\ref{JsymY}). Consequently, the notion of  maximal distinguishability can be generalized only  to particles occupying  different orthogonal modes \cite{Shch2015}.   We will say that  $\bm n$ particles from \textit{different} orthogonal visible  modes are maximally distinguishable when  
\be 
 J_{\hat{\varrho}}(\sigma)= \left\{ \begin{array}{cc} \vare(\sigma), & \sigma\in\mathcal{Y}_{\bm n}\\   0, &  \sigma\notin\mathcal{Y}_{\bm n} \end{array}\right., 
\en{JmaxdistYoung}
where  the  Young subgroup $\mathcal{Y}_{\bm n}$  is defined in Eq. (\ref{YoungSG}).    Particles within a common visible mode have a symmetric label state for bosons and an antisymmetric label state for fermions. Thus a same-mode group of bosons is completely indistinguishable (as in the state of Eq.~(\ref{one-mode})), whereas a same-mode group of fermions emulates bosons in the sense of Corollary~1.

For a fixed occupation vector, departures from Eq.~(\ref{JmaxdistYoung}) produce off-diagonal matrix elements in this occupation basis resulting in multi-particle coherences in  interference experiments.

 \subsection{Coherent superpositions and convex mixtures}
 \label{sec3D}
 
Let us now    use the   indistinguishability  theory to analyze coherent superpositions and convex mixtures of states given by Eq.~(\ref{Psi_nSQ}).  Specifically, we are interested   in the  generalization of    the  label-state indistinguishability theory   to  superpositions and  convex mixtures of visible states. 
 
Before we embark on our analysis,  it is important to distinguish the origin of  a convex mixture of visible states. First of all, if two complete $n$-particle states are mixed,   
\be
\hat{\varrho}_{mix} =  p \hat{\varrho}^{(1)} + (1-p) \hat{\varrho}^{(2)}, \quad 0< p< 1,
\en{mixComplete} 
the visible state is the corresponding convex mixture by linearity of the partial trace. The output probabilities are also convex mixtures, so no interference terms between the components appear.  Thus, for  an incoherent mixture of complete states, no cross-state multi-particle interference  enters the output probabilities. Accordingly,  the   mixture of visible states  taken  alone does not specify a unique assignment of the cross-state  indistinguishability.  In contrast, a cross-state indistinguishability function can be  assigned to  complete states in a coherent superposition, as we show below.

A coherent superposition of complete states can also result in  the  visible state being    a convex mixture of the corresponding visible states.  As an example, consider a coherent superposition of two $n$-particle states in the form of  Eq.~(\ref{Psi_nSQ}), in general, over different sets of orthogonal modes, corresponding to occupation vectors $\bm{n}^{(1)}$ and $\bm{n}^{(2)}$ of  the same basis in $\mathcal{H}_{(vis)}$:
 \be	
 |\Psi_{sup}\rangle = a_1 |\Psi_{\bm{n}^{(1)}}\rangle + a_2 |\Psi_{\bm{n}^{(2)}}\rangle , 
\en{SupPsi}
 with $a_{i}\in  \mathbb{C}$, $i=1,2$, and  $|\Psi_{\bm{n}^{(i)}}\rangle $ being given by Eq. (\ref{Psi_nSQ}) with $\bm{n}$ replaced by $\bm{n}^{(i)}$, $k_\alpha$  by $k_\alpha^{(i)}$, and $C$ by $C^{(i)}$.  The   scalars $a_1,a_2$  are related by the normalization condition, which depends, in general, on the inner product $\langle \Psi_{\bm{n}^{(1)}}|\Psi_{\bm{n}^{(2)}}\rangle$. The visible part of the superposition state of Eq.~(\ref{SupPsi})  reads
 \begin{eqnarray}
 \hat{\varrho}_{(vis)} &= &|a_1|^2  \hat{\varrho}^{(1)}_{(vis)} + |a_2|^2  \hat{\varrho}^{(2)}_{(vis)} \nonumber\\
 &+& a_1^*a_2\mathrm{Tr}_{(int)}\left(|\Psi_{\bm{n}^{(2)}}\rangle \langle \Psi_{\bm{n}^{(1)}}|\right)\nonumber\\
 &+& a_2^*a_1\mathrm{Tr}_{(int)}\left(|\Psi_{\bm{n}^{(1)}}\rangle \langle \Psi_{\bm{n}^{(2)}}|\right),
  \label{SupVis}\end{eqnarray}
 where $\hat{\varrho}^{(i)}_{(vis)} := \mathrm{Tr}_{(int)}\left( |\Psi_{\bm{n}^{(i)}}\rangle \langle \Psi_{\bm{n}^{(i)}}|\right)$, i.e.,    the state of 
 Eq. (\ref{vis_state}) with the  above substitutions. The   cross-terms in Eq.~(\ref{SupVis}) are  obtained by taking the trace over the internal Hilbert space, and thus have a similar form to that of the  visible state in Eq.~(\ref{vis_state}). We get, for instance, 
 \begin{eqnarray}
\label{vis_state21}
&&  \mathrm{Tr}_{(int)}\left(|\Psi_{\bm{n}^{(2)}}\rangle \langle \Psi_{\bm{n}^{(1)}}|\right) \nonumber\\
&& = \!\frac{1}{n!} \sum_{\sigma,\pi\in S_n} \vare(\pi\sigma)  J_{21}(\pi\sigma^{-1})   \frac{\bigotimes\limits_{\alpha=1}^n | k^{(2)}_{\sigma(\alpha)}\rangle\langle k^{(1)}_{\pi(\alpha)}|}{\sqrt{\bm{n}^{(2)}! \bm{n}^{(1)}!}},\nonumber\\
 \end{eqnarray}  
 where $k_1^{(i)}\le \ldots\le  k_n^{(i)}$ is the  sequence of modes corresponding to the occupation vector $\bm{n}^{(i)}$ and we have introduced the cross-indistinguishability function. The latter is an element  of a matrix of indistinguishability:
 \be
 J_{ij}(\sigma):=\mathrm{Tr}\left(\hat{P}_\sigma |\Psi^{(l)}_{\bm{n}^{(i)}}\rangle \langle \Psi^{(l)}_{\bm{n}^{(j)}}|\right),\quad i,j\in \{1,2\},
 \en{indMatJ}
 with the corresponding label states $ |\Psi^{(l)}_{\bm{n}^{(i)}}\rangle$ given by Eq.~(\ref{Psi_nlab}) with the obvious substitutions.  For instance, $J_{\hat{\varrho}^{(i)}} = J_{ii}$ in the new nomenclature. There are two cases. 
 \medskip
 
 \noindent\textit{Case I.}  Superposition of two states with coinciding mode occupations $\bm{n}^{(i)} = \bm{n}$. We  can combine the four terms in Eq.~(\ref{SupVis}) by introducing the total indistinguishability function as follows
 \be
 J_{sup}(\sigma):= \sum_{i,j=1}^2 a_ia^*_j J_{ij}(\sigma).
 \en{Jtot}  
We then obtain a visible state of the form of Eq.~(\ref{vis_state}) with the label state given by a superposition with the same coefficients as the total state in Eq.~(\ref{SupPsi}):
 \be
 |\Psi^{(l)}_{sup}\rangle := a_1 |\Psi^{(l)}_{\bm{n}^{(1)}}\rangle + a_2|\Psi^{(l)}_{\bm{n}^{(2)}}\rangle.
 \en{SupLabel}
 Indeed, one may  take into account that in this  case the superposition  state Eq.~(\ref{SupPsi}) can be recast as the state in Eq.~(\ref{Psi_nSQ}) for the following coefficients $C^{(sup)}_{j_1,\ldots,j_n} =  a_1 C^{(1)}_{j_1,\ldots,j_n}+a_2 C^{(2)}_{j_1,\ldots,j_n}$.     
\medskip

\noindent\textit{Case II.} Superposition of two states with different mode occupations $\bm{n}^{(2)} \ne  \bm{n}^{(1)}$.  The   form of the state crucially depends on the \textit{mutual} indistinguishability properties described by the indistinguishability matrix of Eq.~(\ref{indMatJ}). In general, the elements $J_{21}(\sigma)= J^*_{12}(\sigma^{-1})$  are not  identically zero and   the state  in Eq.~(\ref{SupVis})  contains cross-coherence terms.   The combined visible state becomes a  convex mixture  of the individual visible states only  under  vanishing cross-indistinguishability,
\be
 J_{21}(\sigma)\equiv 0\Rightarrow \hat{\varrho}_{(vis)} = |a_1|^2  \hat{\varrho}^{(1)}_{(vis)} + |a_2|^2  \hat{\varrho}^{(2)}_{(vis)}.
\en{convex_vis}
 
The above analysis was based on states expandable as superpositions of states of the form in Eq.~(\ref{Psi_nSQ}), with definite occupation vectors over a common basis of visible modes. One can  extend Eq.~(\ref{convex_vis})  to superpositions over  different bases of visible modes, as it depends only on the corresponding  label states.

As an application of the above analysis, consider a coherent superposition of complete states of maximally distinguishable particles with identically vanishing cross-indistinguishability functions. Invariably, in this case the visible mixture describes maximally distinguishable particles  in the interference sense defined in Section~\ref{sec3C}. For two components with distinct occupied modes in their respective orthogonal bases, the conditions are
\be
 J_{11}(\sigma)=J_{22}(\sigma)=\delta_{\sigma,e},
 \qquad J_{21}(\sigma)=J_{12}(\sigma)=0.
\en{JmaxdistSuperposition}
They imply, for an arbitrary linear unitary interferometer (acting on the visible modes),
\be
 p_{\bm m}^{(sup)} 
 =\sum_{i=1}^2 |a_i|^2 p_{\bm m;i}^{(d)},
 \qquad \sum_{i=1}^2|a_i|^2=1,
\en{maxdistSuperpositionProb}
where $p_{\bm m;i}^{(d)}$ is the convex mixture of the product of    single-particle probabilities  from the $i$th term,   corresponding to  maximally distinguishable particles (the maximally distinguishable case of Eqs.~(\ref{prob_{nm}})-(\ref{Jmaxdist})). There is no  exchange-interference term   within either component, and, similarly,  no cross-state interference  between them.    Such a state therefore is a state of   maximally distinguishable particles.

 The cross-state condition in Eq.~(\ref{JmaxdistSuperposition}) concerns complete states in a coherent superposition. An incoherent mixture of complete states corresponding to maximally distinguishable particles   has  a similar  probability law to that of Eq.~(\ref{maxdistSuperpositionProb}) by linearity.  In such a  case there is no definite cross-state indistinguishability, which   is not  needed to account for interference experiments with such mixtures.

The   maximal distinguishability defined  via the indistinguishability function turns out to be  a convex property, where the latter applies.    In contrast, the definition based on the visible state of maximal distinguishability in Ref.~\cite{DistNew}  is not convex.   The same  mixture can  satisfy the present   criterion and fail the spectral criterion, where both definitions are based on dynamical invariants.    The  distinctive  feature of the present approach lies in the fact that one cannot assign a definite indistinguishability to identical particles by completely ignoring their internal subspace. 

In Section~\ref{sec4} we also  characterize the  maximal distinguishability by the generalized projective measures, as a byproduct of our  discussion of  indistinguishability functions that are also class functions (i.e., depending only on the cycle structure of a permutation)  on the symmetric group. Moreover, we show  that the  indistinguishability function accounts for  the  matrix-valued  symmetry spectrum of the visible state in the general case.

 \section{Indistinguishability and generalized symmetries}
 \label{sec4}

 Recall that  irreducible characters $\chi_\lambda(\sigma)$ of the symmetric group $S_n$  are  associated with (ordered) integer partitions $\lambda\vdash n$ (Young diagrams), defined as follows  
 \be 
 \lambda=(\lambda_1,\ldots,\lambda_r),\quad    \lambda_1+\ldots +\lambda_r=n,  
 \en{part_lambda}
and  (by convention) $\lambda_i\ge \lambda_{i+1}$.  The    irreducible characters $\chi_\lambda(\sigma)$  are  positive semidefinite (in the sense of Eq.~(\ref{psdJ}))  class functions on $S_n$, i.e., they satisfy   
  \be
  F(\tau^{-1}\sigma\tau)=F(\sigma), \quad \forall \sigma,\tau\in S_n. 
  \en{classF}
They are also real-valued and form an orthonormal basis of the class functions with respect to    the following inner product on functions on $S_n$
 \be  \left(f,g\right) := \frac{1}{n!}\sum_\sigma f^*(\sigma)g(\sigma). 
 \en{iprod}
    
 The  important observation is that  the normalized  irreducible characters $\chi_\lambda(\sigma)/\chi_\lambda(e)$ (recall that $e$ is the identity in $S_n$)  
  are valid indistinguishability functions.  The normalization here is by  the dimension of the associated irreducible representation 
 \be
d_\lambda:=\chi_\lambda(e).
\en{dchilambda}
 For example, bosons and fermions correspond to the following  one-dimensional  characters
 \be 
 \chi_{(n)}(\sigma)\equiv 1,\quad \chi_{(1,\ldots,1)}(\sigma)= \mathrm{sgn}(\sigma),
 \en{bfchar}
 respectively.   The  projective measures in   Eq.~(\ref{DJ}) are    projections    on these two  characters. 
 
 \subsection{Permutation-invariant   labels}
\label{sec4A}

The indistinguishability function is a class function on $S_n$ when the  internal state can serve as the label state (recall that the map from label states  to indistinguishability functions is many-to-one).  Indeed,   for a class function $J_{\hat{\varrho}}(\sigma) = J_{\hat{\varrho}}(\pi^{-1}\sigma\pi)$ for all $\sigma,\pi\in S_n$,  we can perform the   averaging: 
 \be
 J_{\hat{\varrho}}(\sigma) = \frac{1}{n!}\sum_{\pi\in S_n} J_{\hat{\varrho}}(\pi^{-1}\sigma\pi),
 \en{permAvJ}
 which implies   that there is a label state satisfying $\hat{P}_\pi\hat{\varrho}^{(l)}\hat{P}^\dag_\pi = \hat{\varrho}^{(l)}$, for all $\pi\in S_n$.  Since we  need only the indistinguishability function and not the label state itself (see also the discussion in Section \ref{sec4C} below),  we can   set $ \hat{\varrho}^{(l)}=\hat{\varrho}_{(int)}$.

 For example, if $n$ identical particles occupy distinct orthogonal visible modes and are prepared independently in the same mixed single-particle label state   $\hat{\rho}$  (see Appendix~\ref{appB} for more details),  we have    $\hat{\varrho}^{(l)} = \hat{\rho}^{\otimes n}$ with the class-function indistinguishability.  
 This case is realized, for example, by independent single-particle emissions from a stable source with no memory between emissions (e.g., in the standard quantum optics nomenclature the label state $\hat{\rho}$ is called simply a state of a photon).  
Let 
  \be
\hat{\rho}=\sum_j  q_j |f_j\rangle\langle f_j|,\quad q_j>0, \quad \sum_j q_j=1.
\en{state_rho} 
 The corresponding  indistinguishability function  reads   \cite{Shch2014,Shch2015}
\begin{eqnarray}
&& J_{\hat{\rho}^{\otimes n}}(\sigma) = \operatorname{Tr}(\hat{P}_\sigma \hat{\rho}^{\otimes n})= \sum_{j_1,\ldots,j_n}   \prod_{\alpha=1}^n q_{j_\alpha}  \delta_{j_\alpha,j_{\sigma(\alpha)}} \nonumber\\
&&  =   \prod_{k=1}^{n} \Bigl(\sum_j q_j^k\Bigr)^{C_k(\sigma)} = \prod_{k=1}^{n} \left(\mathrm{Tr}\hat\rho^k\right)^{C_k(\sigma)}  ,
\label{J_tensor}
\end{eqnarray} 
 where    $C_k(\sigma)$ denotes the number of cycles of length $k$  in the cycle decomposition of  $\sigma$, i.e., $J_{\hat{\rho}^{\otimes n}}$ depends only on the partition type $\lambda$ of $\sigma$, where $C_k(\sigma)$ is the number of elements $\lambda_i=k$ in the partition.

Any convex mixture of the $n$th tensor powers of single-particle label states   is also a  permutation-invariant label state of $n$ identical particles.   The converse is not true:  there are permutation-invariant label states   which  cannot be expanded  as a  convex mixture  of  the $n$-th tensor powers of single-particle label states.   An explicit example  is as follows.   
Consider the   states  $|j,s\rangle\in  \mathcal{H}_{(int)}$ such that $|j,s\rangle\equiv |s\rangle\otimes |j\rangle$, with $s\in\{\pm\}$ and $j=1,2,3,\ldots$.  Physically, the $s$-label can be spin (polarization), while the $j$-label can be a  basis state of temporal shapes.  Let $p_{j,s}\geq0$, $\sum_{j,s}p_{j,s}=1$, and   $p_j=p_{j,+}+p_{j,-}$.  Introduce two commuting single-particle label states  as follows 
 \[
 \hat{\rho}_1 =\sum_{j,s}\frac{p_j}{2}|j,s\rangle\langle j,s|,
 \quad
\hat{\rho}_2=\sum_{j,s}p_{j,s}|j,s\rangle\langle j,s|.
\]
   For $x>0$, consider the trace-normalized signed combination
\be
\hat{A} (x):=(1+x) \hat{\rho}_1^{\otimes n}-x\hat{\rho}_2^{\otimes n},
 \en{Asigned}
The eigenvalue of  $\hat{A}(x)$ corresponding to the    eigenvector  $\bigotimes_{\alpha=1}^n |j_\alpha,s_\alpha\rangle$  is
\[
 (1+x)\prod_{k=1}^{n}\frac{p_{j_k}}{2} -  x\prod_{k=1}^{n}p_{j_k,s_k}.
\]
Introduce   
\[
 R=\sup_{\substack{(j,s); p_j>0}}\frac{2p_{j,s}}{p_j},
\]
  where   $1\leq R\leq2$. The operator $\hat{A}(x)$ in Eq. (\ref{Asigned})  is  positive semidefinite for  $R>1$ and  
 \be
 x\leq\frac{1}{R^n-1}. 
\en{xbound}
Furthermore, $\hat A(x)$ cannot be a convex mixture of $n$th tensor powers for $n\ge2$, $R>1$, and $x>0$. To see this, choose a bounded Hermitian single-particle operator $B$ such that $b_1=\mathrm{Tr}(B\hat\rho_1)\ne b_2=\mathrm{Tr}(B\hat\rho_2)$. The one- and two-particle marginals $A_1$ and $A_2$ satisfy
\[
 \mathrm{Tr}[(B\otimes B)A_2]-[\mathrm{Tr}(BA_1)]^2
 =-x(1+x)(b_1-b_2)^2<0.
\]
For a mixture $\int\hat\rho^{\otimes n}\,d\mu(\hat\rho)$, the same quantity is the variance of $\mathrm{Tr}(B\hat\rho)$ under $\mu$ and is nonnegative. This proves the claim.

  The complete characterization of permutation-invariant indistinguishability   is supplied by the character theory of the symmetric group. Indeed,   using  Proposition 1 of Section \ref{sec3A} one can prove the following proposition (see   Appendix \ref{appC} for details).
\begin{proposition}
Every positive semidefinite class function $J(\sigma)$ on $S_n$, satisfying also $J(e)=1$, can be expanded as the following convex sum  
\be
J(\sigma) = \sum_{\lambda\vdash n}\mathcal{D}^{(\lambda)}  \frac{\chi_\lambda(\sigma)}{\chi_\lambda(e)},\quad \mathcal{D}^{(\lambda)} \ge 0,\quad \sum_\lambda \mathcal{D}^{(\lambda)} =1.
\en{Jnew}
 \end{proposition}
Therefore, the    indistinguishability functions corresponding to permutation-invariant labels form  a convex compact set with the extreme points  given by normalized  irreducible characters of the symmetric group. 

By the mutual  orthogonality of $\chi_\lambda$ with respect to the inner product in Eq.~(\ref{iprod}), we get  from Eq.~(\ref{Jnew})
\be
 \mathcal{D}^{(\lambda)}=\frac{\chi_\lambda(e)}{n!}\sum_\sigma \chi_\lambda(\sigma)J(\sigma),
 \en{Dlambda}
 for instance   $\mathcal{D}^{((n))}=\mathcal{D}$ and  $\mathcal{D}^{((1,\ldots,1))}=\widetilde{\mathcal{D}}$ by Eq.~(\ref{DJ}).

The Young-subgroup condition in Eq.~(\ref{JsymY}) must also hold. In particular, if any mode is multiply occupied and $J$ is a class function, then $J$ is forced to be $1$ for bosons and $\mathrm{sgn}$ for fermions: the Young subgroup contains a transposition, all transpositions are in the same cycle class, and an expectation value of $+1$ or $-1$ for each transposition implies the above  conclusion. Therefore, nontrivial class-function partial indistinguishability   examples occur for  single particles.

For an arbitrary  state $\hat{\varrho}$ of $n$ identical particles  there  is a generalization of Theorem 1 and Corollary 1  (see Appendix~\ref{appC}).   
\begin{theorem} We have 
  \be	
\mathcal D^{(\lambda)} =  \mathrm{Tr}\!\left(\hat{S}^{(\lambda)}\hat{\varrho}_{(int)}\right)= \mathrm{Tr}\!\left(\hat{S}^{(\lambda^\prime)}\hat{\varrho}_{(vis)}\right) ,
\en{D_lambda}
where $\hat{S}^{(\lambda)}$ belongs to the complete set of mutually orthogonal  projectors onto the  $\lambda$-symmetry sectors
\be
\hat{S}^{(\lambda)} := \frac{\chi_\lambda(e)}{n!} \sum_{\sigma \in S_n} \chi_\lambda(\sigma)\hat{P}_\sigma,
\en{S_lambda}
 $\lambda^\prime = \lambda$ for bosons, and  $\lambda^\prime=\lambda^T$  for fermions, with $\lambda^T$ being  the transposed partition (for  the transposed Young diagram) corresponding to $\chi_{\lambda^T}(\sigma) = \mathrm{sgn}(\sigma)\chi_\lambda(\sigma)$. 

Moreover, the set    $\{\mathcal{D}^{(\lambda^\prime)}\}$ has the physical meaning of  the  probability distribution  on   the  symmetry spectrum  $\{\hat{\varrho}_{(vis)}^{(\lambda)}\}$. We have 
\be
 \hat{\varrho} _{(vis)} =  \bigoplus_{\lambda\vdash n} \hat{S}^{(\lambda)} \hat{\varrho}_{(vis)}\hat{S}^{(\lambda)} =\bigoplus_{\lambda\vdash n}  \mathcal{D}^{(\lambda^\prime)} \hat{\varrho}^{(\lambda)} _{(vis)},
\en{vis_st_lambda}
where $\hat{\varrho}^{(\lambda)}_{(vis)}:=\hat{S}^{(\lambda)}\hat{\varrho}_{(vis)}\hat{S}^{(\lambda)}/\mathcal{D}^{(\lambda^\prime)}$ for $\mathcal D^{(\lambda^\prime)}>0$. (Terms with zero weight are omitted from the normalized decomposition.)
\end{theorem}

Therefore, the  projective  measures on the generalized symmetry sectors completely characterize class-function indistinguishability. By the reconstruction formula Eq. (\ref{vis_state}), they also  determine the visible state  for a given occupation vector  over orthogonal  visible modes.

 \subsection{Characterization of maximal distinguishability  }
 \label{sec4B}

From  Section \ref{sec3} we know that the  maximal distinguishability corresponds to the indistinguishability function  Eq.~(\ref{Jmaxdist})  
and  occurs for single particles occupying distinct visible modes.  Within the class-function family, Theorem~3 provides a set of conditions singling out the maximally distinguishable states.

For the unrestricted examples in this subsection, assume $M\ge n$ and $\dim\mathcal H_{(int)}\ge n$. Before we present the result for the general case,    let us  analyze     the simplest non-trivial case of  $n= 3$ particles,  whose indistinguishability function is a class function on $S_3$. In this case  Eq.~(\ref{Jnew})  results in    a  one-parameter family of non-diagonal visible states in the form of Eq. (\ref{vis_state}),  with $\mathcal{D} =1/n!=1/6$,    defined  by     $\widetilde{\mathcal{D}}$ (a free parameter  due to the existence of yet  another partition $\lambda = (2,1)$). For all $\widetilde{\mathcal{D}}\ne 1/6$ we get  $J(\sigma)\ne \delta_{\sigma,e}$, thereby   resolving a problem posed in   Ref.~\cite{DistNew} (see the details in  Appendix C).   

Furthermore,  for $n\ge 4$  the expansion of Eq.~(\ref{Jnew})  contains a family of visible states  in the form of Eq. (\ref{vis_state})  which satisfy both conditions in Eq.~(\ref{Dmaxdist}) and have  $p_n-3$ free non-negative  parameters, where $p_n$ is the number of partitions Eq.~(\ref{part_lambda}) in $S_n$, e.g.,  $p_4 = 5$. Thus  Eq.~(\ref{Dmaxdist}), previously proposed to  identify the maximally distinguishable case  in Ref. ~\cite{WeylD}, also does not single out the maximally distinguishable states.

 The above examples show that one has to consider all the generalized symmetry sectors to characterize the maximally distinguishable case. For class-function indistinguishability, Theorem~3 characterizes maximal distinguishability as a special probability distribution $\{ \mathcal{D}^{(\lambda^\prime)}_d\}$ on the symmetry sectors. Indeed,  it is known that 
 \be
 \chi_R(\sigma):= n!\delta_{\sigma,e}
 \en{regchi}
  is   the character of the so-called regular representation of the symmetric group  $S_n$,  acting on a linear vector  space of dimension $d_R=n!$  as    permutations of the basis states. The   well-known decomposition of the regular character  into the irreducible characters  immediately gives
\be
J^{(d)}(\sigma) =  \frac{1}{n!}\sum_{\lambda\vdash n} \chi_\lambda(e) \chi_\lambda(\sigma)= \sum_{\lambda\vdash n} \mathcal{D}^{(\lambda)}_d  \frac{ \chi_\lambda(\sigma)}{\chi_\lambda(e)},
\en{MdJ}
with
\be
\mathcal{D}^{(\lambda)}_d :=\frac{\chi^2_{\lambda}(e)}{n!}=\frac{d^2_{\lambda}}{n!}= \mathcal{D}^{(\lambda^\prime)}_d,
\en{Planch}
where we use the fact that  $ \chi_{\lambda^\prime}(e) = \vare(e)\chi_\lambda(e) = \chi_\lambda(e)=d_\lambda$.   Eq.~(\ref{Planch})  is the  well-known Plancherel distribution over the symmetry sectors $\lambda\vdash n$.

For  a general indistinguishability function, which is not a class function,  the weights in  Eq.~(\ref{Planch})  are   necessary  but not sufficient for  maximal distinguishability, as we show in  the next section.

 
  \subsection{ Matrix-valued symmetry spectrum and indistinguishability function }
\label{sec4C}

The characterization by the projective measures
$\{\mathcal D^{(\lambda)}\}$  is  incomplete if the  indistinguishability function is not a class function.   The additional information, beyond the permutation-invariant average of  Eq.~(\ref{permAvJ}), contained in a general indistinguishability function   accounts for the matrix-valued symmetry  (Fourier) spectrum  \cite{GB2026}.  Below, we discuss the equivalence of  the present approach based on  indistinguishability function to the Fourier spectrum  approach  Ref.~\cite{GB2026} and point  out  the advantages of the former.  

Let $D^\lambda(\sigma)$ be a real orthogonal irreducible representation of $S_n$ associated with $\lambda\vdash n$. Such a choice is always possible for $S_n$ and does not restrict $J$ to be real-valued (in contrast to class-function indistinguishability in Eq.~(\ref{Jnew})).  For an arbitrary (positive semidefinite) indistinguishability function $J(\sigma)$, normalized by $J(e)=1$, introduce
\be
\hat A_\lambda :=
\frac{d_\lambda}{n!}
\sum_{\sigma\in S_n}
J(\sigma)D^\lambda(\sigma^{-1}).
\en{Alambda}
The positivity of $J$ on the group implies
\[
\hat A_\lambda\geq0 ,
\]
for every $\lambda\vdash n$.  The Fourier inversion formula on the finite group $S_n$ gives
\be
J(\sigma)
=
\sum_{\lambda\vdash n}
\operatorname{Tr}\!\left[
\hat A_\lambda D^\lambda(\sigma)
\right].
\en{JFourier}
The normalization $J(e)=1$ becomes
\be
\sum_{\lambda\vdash n}\operatorname{Tr}\hat A_\lambda=1.
\en{Anorm}
Therefore, the  matrix symmetry spectrum 
\be
\{\hat A_\lambda:\lambda\vdash n\}
\en{symspec}
is necessary for a complete characterization of a general indistinguishability function. 

The trace of $\hat A_\lambda$ coincides exactly with the projective measure introduced above.  Indeed, since
$\operatorname{Tr}D^\lambda(\sigma^{-1})
=\chi_\lambda(\sigma)$ for $S_n$, Eq.~(\ref{Alambda}) gives
\be
\operatorname{Tr}\hat A_\lambda
=
\frac{d_\lambda}{n!}
\sum_{\sigma\in S_n}
\chi_\lambda(\sigma)J(\sigma)
= \mathcal D^{(\lambda)},
\en{traceAlambda}
i.e.,  the projective symmetry spectrum
$\{\mathcal D^{(\lambda)}\}$ is   the  trace  of the more general matrix-valued spectrum $\{\hat A_\lambda\}$.
The   averaging as in Eq.~(\ref{permAvJ}) acts   in every irreducible sector by replacing the positive semidefinite Hermitian matrices   as follows
\be
\hat A_\lambda
\longmapsto
\frac{\operatorname{Tr}\hat A_\lambda}{d_\lambda}\hat I_{d_\lambda}.
\en{Atwirl}
Consequently,  for class-function indistinguishability,
\[
J(\tau^{-1}\sigma\tau)=J(\sigma),
\qquad
\forall\,\sigma,\tau\in S_n,
\]
Eq.~(\ref{Alambda}) implies
\[
D^\lambda(\tau)\hat A_\lambda
D^\lambda(\tau)^{-1}
=
\hat A_\lambda,
\qquad
\forall\,\tau\in S_n.
\]
By Schur's lemma,
\be
\hat A_\lambda
=
\frac{\mathcal D^{(\lambda)}}{d_\lambda}
\hat I_{d_\lambda}.
\en{Aclass}
Substitution into Eq.~(\ref{JFourier}) immediately recovers the character expansion
\[
J(\sigma)
=
\sum_{\lambda\vdash n}
\mathcal D^{(\lambda)}
\frac{\chi_\lambda(\sigma)}{d_\lambda},
\]
  confirming  that the projective measures completely characterize  partial indistinguishability   when the indistinguishability function is a class function.

The physical meaning of the matrices $\hat A_\lambda$ is particularly clear for particles occupying $n$ distinct orthogonal visible modes.  The states
\[
|\sigma\rangle
=
\hat P_\sigma
\bigotimes_{k=1}^n|k\rangle,
\qquad
\sigma\in S_n,
\]
span an $n!$-dimensional space carrying the regular representation,
\[
\mathcal V_{\rm reg}
\simeq
\bigoplus_{\lambda\vdash n}
V_\lambda\otimes\mathbb C^{d_\lambda}.
\]
The first factor $V_\lambda$ carries the irreducible representation $D^\lambda$, whereas the second factor accounts for its multiplicity $d_\lambda$ in the regular representation.

For bosons, the visible state of Eq.~(\ref{vis_state}) in this space reads
\[
\hat\varrho_{(vis)}
=
\frac{1}{n!}
\sum_{\sigma,\pi\in S_n}
J(\pi^{-1}\sigma) |\sigma\rangle\langle\pi|.
\]
With the Fourier convention
\[
 |\sigma\rangle\longmapsto
 \bigoplus_{\lambda\vdash n}\sqrt{\frac{d_\lambda}{n!}}
 \sum_{a,b=1}^{d_\lambda}D^\lambda(\sigma)_{ab}
 |a\rangle\otimes|b\rangle,
\]
Schur orthogonality gives the block form
\be
\hat\varrho_{(vis)}
=
\bigoplus_{\lambda\vdash n}
\frac{\hat I_{d_\lambda}}{d_\lambda}
\otimes
\hat A_\lambda.
\en{rhoAlambda}
In particular, the maximal distinguishability condition Eq.~(\ref{Jmaxdist})  becomes equivalent to   
\[
 \hat A_\lambda=\frac{d_\lambda}{n!}\hat I_{d_\lambda},
 \quad\forall \lambda,
\]
which is stronger than the condition in Eq.~(\ref{Jmaxdist}) for class-function indistinguishability.   

  The first factor in Eq.~(\ref{rhoAlambda}) is fixed by the permutation invariance of the visible state, whereas $\hat A_\lambda$ determines its structure in the multiplicity space.  Equation~(\ref{rhoAlambda}) immediately gives
\[
\operatorname{Tr}
\left(
\hat S^{(\lambda)}
\hat\varrho_{(vis)}
\right)
=
\operatorname{Tr}\hat A_\lambda
=
\mathcal D^{(\lambda)}.
\]
Thus the projective measure $\mathcal D^{(\lambda)}$ specifies only the total weight of the visible state in the symmetry sector $\lambda$, while the normalized matrix
\be
\hat\rho_\lambda^{(m)}
:=
\frac{\hat A_\lambda}{\mathcal D^{(\lambda)}},
\qquad
\mathcal D^{(\lambda)}>0,
\en{rhomult}
specifies the state within the corresponding multiplicity degrees of freedom.

For class-function indistinguishability,
Eq.~(\ref{Aclass}) gives
\[
\hat\rho_\lambda^{(m)}
=
\frac{\hat I_{d_\lambda}}{d_\lambda},
\]
so no information remains inside the multiplicity space and the scalar probabilities $\{\mathcal D^{(\lambda)}\}$ provide a complete description.  

For fermions the above construction also applies after multiplication of the indistinguishability function by the signature, as in the uniform function $\Lambda(\sigma)=\varepsilon(\sigma)J(\sigma)$ introduced in Section~\ref{sec3A}. In this case, the label symmetry sector $\lambda$ is associated with the visible sector $\lambda^T$, in accordance with Theorem~3.

 The indistinguishability function and the matrix-valued symmetry spectrum  also agree on  the free parameter count. Consider first $n$ particles occupying $n$ distinct visible modes, for which the Young subgroup $\mathcal Y_{\bm n}$ Eq.~(\ref{YoungSG}) is trivial. A Hermitian  matrix $\hat A_\lambda$ of size $d_\lambda$ contains $d_\lambda^2$ free real parameters. Since the   dimensions  of the irreducible  representations  satisfy
\[
\sum_{\lambda\vdash n}d_\lambda^2=n!,
\]
the collection $\{\hat A_\lambda\}$ contains $n!$ real parameters before normalization. Equation~(\ref{Anorm}) removes one parameter, leaving $n!-1$. On the other hand, the  indistinguishability function, being a positive semidefinite function on $S_n$,     satisfies the  relation
\[
J(\sigma^{-1})=J(\sigma)^*	
\]
(see  Eq.~(\ref{distJ})) and, hence,   contains $n!$ real parameters, while  the normalization $J(e)=1$  leaves $n!-1$ free parameters. Therefore, both  $J(\sigma)$ Eq.~(\ref{JFourier}) and the associated  matrix-valued symmetry spectrum Eq.~(\ref{symspec})  carry exactly the same amount of information.

The parameter count extends to an occupation vector $\bm n$ containing repetitions. Define
\begin{align}
 Q_\lambda^{(\varepsilon)}
 &=\frac{1}{\bm n!}\sum_{\pi\in\mathcal Y_{\bm n}}
       \vare(\pi)D^\lambda(\pi),\nonumber\\
 k_\lambda^{(\varepsilon)}
 &=\mathrm{Tr}\,Q_\lambda^{(\varepsilon)}
 =\frac{1}{\bm n!}\sum_{\pi\in\mathcal Y_{\bm n}}
       \vare(\pi)\chi_\lambda(\pi).
 \label{young_rank}
\end{align}
The Young-subgroup condition gives
$\hat A_\lambda=Q_\lambda^{(\varepsilon)}\hat A_\lambda Q_\lambda^{(\varepsilon)}$.
Thus, when the internal Hilbert space contains all the required sectors, the number of free real parameters is
\be
 \sum_{\lambda\vdash n}\bigl(k_\lambda^{(\varepsilon)}\bigr)^2-1.
 \en{count_repetitions}
The same count follows directly from $J$, whose independent values are constrained by the left and right Young-subgroup actions and its Hermitian symmetry. If the internal dimension is $d<n$, the sum is restricted to partitions with at most $d$ rows. For distinct occupied modes, the Young subgroup is trivial and $k_\lambda^{(\varepsilon)}=d_\lambda$, recovering $n!-1$.

The equivalence established above between the indistinguishability function and the matrix-valued symmetry spectrum allows us to fully  appreciate the central role of the former: $J_{\hat{\varrho}}(\sigma)$   contains all information on partial indistinguishability required to determine the visible state Eq.~(\ref{vis_state}).   The matrix-valued symmetry spectrum $\{\hat A_\lambda\}$ does not introduce additional information; by Eqs.~(\ref{Alambda}) and (\ref{JFourier}) it is related to $J_{\hat{\varrho}}(\sigma)$   by an invertible  transformation. The role of the matrix-valued spectrum is  to resolve the information contained in $J$ into generalized symmetry sectors.  The observable quantities, e.g., probabilities,   can be derived directly from  $J(\sigma)$, without   constructing the irreducible representations to carry  out the Schur--Weyl decomposition.

The important practical advantage of the indistinguishability function Eq.~(\ref{distJ})  also stems from its  independence of the  evolution of the visible part. The indistinguishability function 
is  interferometer-independent. Thus, once it  has been determined for a given preparation, the same indistinguishability function can be used for any linear interferometer by changing only the matrix elements of $\hat U$ in Eq.~(\ref{prob_{nm}}), in contrast, e.g.,  to the approach based on the Schur--Weyl decomposition \cite{WeylD,DistNew}.    This separation between the invariant information carried by the preparation and the variable visible evolution is one of the principal practical advantages of the indistinguishability-function formalism.
The equivalent   matrix-valued Fourier spectrum Eq.~(\ref{symspec})  is likewise independent of the interferometer, since it is completely determined by $J(\sigma)$ through Eq.~(\ref{Alambda}). Its use, however, requires the construction of the irreducible representations $D^\lambda$, together with the corresponding decomposition into irreducible and multiplicity spaces. As we have shown, none  of these representation-theoretic constructions is really  necessary for the calculation of  observable quantities.     
 
 
 \section{Hierarchy of   indistinguishability}
 \label{sec5}

 The complement of the projective measure    $\mathcal{D} (\hat{\varrho})$  of Theorem 1 of Section \ref{sec2}   is an upper bound on the total variation distance between the output probability distributions of a linear interferometer for the given and ideal cases. However,   estimating the projective measure of $n$ identical particles in an interference  experiment   is challenging   for  $n\gg 1$.  For example, for  bosons, $\mathcal{D} (\hat{\varrho})$   appears as a quantum-enhancement factor   in the  probability  of boson bunching in a single output mode \cite{Shch2015A,Tch2015}.  However,   the output probabilities are typically  exponentially small in $n$ (at least for $M\ge n$). Indeed,  for  ideally indistinguishable bosons  in a Haar-random unitary linear interferometer  the average probability $p_{\bm{m},\bm{n}}$,    Eq.~(\ref{prob_{nm}}),   becomes  independent of  $\bm{n}$ and $\bm{m}$, thus being given by the reciprocal of the number of distributions of $n$ bosons over  $M$ modes
  \be
\langle p_{\bm{m},\bm{n}}\rangle_{\hat{U}} = \binom{M+n-1}{M-1}^{-1} = \frac{n!}{(M+n-1)\ldots M}.
 \en{pnm_AV} 
 
Below we describe  a hierarchy of upper bounds on the projective indistinguishability measure which allows one to balance the amount of data to be collected against the order of the upper bound on indistinguishability for an  arbitrary number $n$ of identical particles.  The hierarchy is based on the marginal probabilities for $r\le n$ particles in an interference experiment  involving  $n$ identical particles.   
The  higher-order marginal probabilities are intimately related to   collective  multiparticle   phases, manifestations of multiparticle correlation functions  \cite{nphPhases}. Similar approach has been adopted to study averaged higher-order correlations of  many-body states \cite{MBCoh22}. 
 
 \subsection{Marginal   $r$-particle  probabilities }
 \label{sec5A}
 
 To introduce the marginal probabilities we will rely on the following summation identity relating   summation over occupations to  that over  modes. For an arbitrary symmetric function    $f(l_1,\ldots, l_n)$, where $1\le l_\alpha\le M$,  we have 
 \be
 \sum_{\bm{m}}f(l_1,\ldots, l_n) = \sum_{l_1=1}^M \ldots \sum_{l_n=1}^M \frac{\bm{m}!}{n!} f(l_1,\ldots, l_n) 
 \en{sum_ID}
where  $\bm{m} = (m_1,\ldots, m_M)$, $m_j$  being the number of $l_\alpha=j$ in  the sequence $l_1,\ldots, l_n$. 
Applying  the summation identity in  Eq.~(\ref{sum_ID}) to the   output probability  $p_{\bm{m},\bm{n}}$  Eq.~(\ref{prob_{nm}}),  partitioning the output-mode sequence into $r$ and $n-r$ entries together with the corresponding occupation vector $\bm{m} = \bm{m}^\prime + \bm{m}^{\prime\prime}$ and applying the identity Eq.~(\ref{sum_ID}) in reverse, we obtain 
\begin{eqnarray}
&& \sum_{\bm{m}} p_{\bm{m},\bm{n}} = \sum_{l_1=1}^M \ldots \sum_{l_n=1}^M \frac{\bm{m}!}{n!} p_{\bm{m},\bm{n}}\nonumber\\
 && =  \sum_{\bm{m}^\prime} \frac{r!}{\bm{m}^\prime!} \sum_{\bm{m}^{\prime\prime}}\frac{(n-r)!}{\bm{m}^{\prime\prime}!} \frac{\bm{m}!}{n!} p_{\bm{m},\bm{n}} \nonumber\\
 && =  \binom{n}{r}^{-1}\sum_{\bm{m}^{\prime}} \sum_{\bm{m} \supset \bm{m}^{\prime}} \frac{\bm{m}!}{\bm{m}^\prime!(\bm{m}\!-\!\bm{m}^\prime) !}p_{\bm{m},\bm{n}},
 \label{prob_ID}
\end{eqnarray}
 where, in the last line, the inner sum is over all occupations $\bm{m}$ expandable as $\bm{m}  = \bm{m}^{\prime}+ \bm{m}^{\prime\prime}$ with $m^{\prime\prime}_l\ge 0$. Since the  terms in  Eq.~(\ref{prob_ID})  sum to  $1$ by definition of probability,  we can  identify the  term  in the summation over $\bm{m}^{\prime}$, $|\bm{m}^\prime|:=m_1^{\prime}+\ldots m^{\prime}_M=r$,  as the factorial marginal probability for $\bm m^\prime$. Operationally, it can be computed from   detected $n$-particle events as follows 
 \begin{eqnarray}
 \label{Prn}
  p^{(r|n)}_{\bm{m}^{\prime},\bm{n}} :=   \binom{n}{r}^{-1} \sum_{\bm{m} \supset \bm{m}^{\prime}} \frac{\bm{m}!}{\bm{m}^\prime!(\bm{m}\!-\!\bm{m}^\prime) !}p_{\bm{m}, \bm{n}}.
  \end{eqnarray} 
 The factor multiplying $p_{\bm{m},\bm{n}}$ in Eq.~(\ref{Prn}) is the multivariate hypergeometric (Fisher-Yates) weight induced by uniform selection of $r$ particles without replacement. Equivalently, it is the weight of a contingency table with fixed margins,     in our case    $(\bm{m}^\prime, \bm{m}^{\prime\prime})$ (i.e., $M$ rows by two columns) with margins $m^\prime_l + m^{\prime\prime}_l  = m_l$ and $(|\bm{m}^\prime|, | \bm{m}^{\prime\prime}|) = (r,n-r)$. 
 Eq.~(\ref{prob_ID})  guarantees that  
\[
  \sum_{\bm{m}^{\prime}}  p^{(r|n)}_{\bm{m}^{\prime},\bm{n}} = 1,
\]
i.e.,   that  $p^{(r|n)}_{\bm{m}^{\prime},\bm{n}} $ is a valid probability distribution. 

In an experiment the  probability Eq.~(\ref{Prn}) can be estimated  by replacing $p_{\bm{m},\bm{n}}$ with the relative frequency  of $\bm{m}$   obtained in  the particle counting detection.   The sampling cost depends on the marginal order $r$, the number of modes $M$, the measured probability, and the requested precision. For a fixed $r$, this cost can be much smaller than the cost of estimating the full $n$-particle distribution, though it is not independent of $n$ when $M$ grows with $n$.   Indeed, for ideally  indistinguishable bosons   in a Haar-random unitary linear interferometer   the average marginal  probability in Eq.~(\ref{Prn}) is  equal to  the reciprocal of the number of occupation vectors $\bm{m}^\prime$
\be
\langle  p^{(r|n) }_{\bm{m}^{\prime},\bm{n}} \rangle_{\hat{U}} =\binom{M+r-1}{M-1}^{-1}  = \frac{r!}{(M+r-1)\ldots M}.
\en{Prn_AV}

Let us  now introduce the  detection operator $\hat{\Pi}^{(r|n)}_{\bm{m}^\prime}$ whose average  on the visible state  Eq.~(\ref{vis_state}) gives the  probability $p^{(r|n)}_{\bm{m}^{\prime},\bm{n}} $ Eq.~(\ref{Prn}).    From the definition Eq.~(\ref{DetOp})    by utilizing the  summation identity Eq.~(\ref{sum_ID}) we obtain (see details in Appendix \ref{appD})
\begin{eqnarray}
\label{detPirn}
&&\hat{\Pi}^{(r|n)}_{\bm{m}^\prime} :=   \binom{n}{r}^{-1} \sum_{\bm{m} \supset \bm{m}^{\prime}} \frac{\bm{m}!}{\bm{m}^\prime!(\bm{m}\!-\!\bm{m}^\prime) !}\hat{\Pi}_{\bm{m}}\nonumber\\
&&= \binom{n}{r}^{-1} \sum_{\nu\in S_{r|n}} \hat{P}_\nu\left[\hat{\Pi}_{\bm{m}^\prime} \otimes \hat{I}^{\otimes n-r}\right] \hat{P}^\dag_\nu
\end{eqnarray}
where $S_{r|n}:=S_n/(S_r\times S_{n-r})$ is the set of order-preserving $(r,n-r)$-shuffles. These permutations select  $r$   single-particle  Hilbert spaces in the tensor power $\mathcal{H}_{(vis)}^{\otimes n}$  to be acted on by the  $r$-particle detection operator  in  output modes of Eqs.~(\ref{Uab2})-(\ref{Uab1}),   
 \be
\hat{\Pi}_{\bm{m}^\prime}=  \frac{1}{ \bm{m}^\prime!}\sum_{\tau\in S_r} \hat{P}_\tau\left( \bigotimes_{\alpha=1}^r |\tilde{l}_\alpha\rangle\langle \tilde{l}_\alpha | \right) \hat{P}^\dag_\tau
 \en{rPi}
   (here the sequence of   modes $l_1,\ldots, l_r$ corresponds to the occupation vector $\bm{m}^\prime$). 
 
Using the expression for the  detection operator in Eqs.~(\ref{detPirn})-(\ref{rPi}) we can rewrite  the marginal probability   in Eq.~(\ref{Prn})   as an $r$-particle probability for the    marginal $r$-particle  visible state    (see details in Appendix \ref{appD}):
  \be
 p^{(r|n) }_{\bm{m}^{\prime},\bm{n}} = \mathrm{Tr}\left( \hat{\Pi}^{(r|n)}_{\bm{m}^\prime} \hat{\varrho}_{(vis)}\right) =  \mathrm{Tr}\left( \hat{\Pi}_{\bm{m}^\prime} \hat{\varrho}^{(r|n)}_{(vis)}\right),
 \en{probmar_vis}
where  the marginal $r$-particle visible state  reads 
\begin{eqnarray}
  \hat{\varrho}^{(r|n)}_{(vis)}: =   \mathrm{Tr}_{{r+1},\ldots, n}{\hat{\varrho}_{(vis)}}
  =  \binom{n}{r}^{-1} \sum_{\bm{\alpha}} \hat{\varrho}^{(\bm{n}^\prime)}_{(vis)} ,
\label{mar_vis}
 \end{eqnarray} 
 with  $\bm{\alpha}:=\{\alpha_1,\ldots,\alpha_r\}$   denoting a subset of $r$ particles  (chosen from $n$ particles in the input occupation vector $\bm{n}$)  and the $r$-particle visible states   as follows
 \be
\hat{\varrho}^{(\bm{n}^\prime)}_{(vis)}: =\frac{1}{r!\bm{n}^\prime!}\!\!\sum_{\sigma,\pi\in S_r} \vare(\pi\sigma) J_{\bm{n}^\prime} (\pi\sigma^{-1})  \bigotimes\limits_{i=1}^r | k_{\alpha_{\sigma(i)}}\rangle\langle k_{\alpha_{\pi(i)}}|.
 \en{rvis}
 In Eq.~(\ref{rvis})    $\bm{n}^\prime$   is the  occupation vector  of $r$ particles in  input modes $k_{\alpha_1},\ldots, k_{\alpha_r}$  and  $J_{\bm{n}^\prime}(\sigma)$  denotes  their   indistinguishability function.
Explicitly, for  $\nu \in  S_{r|n}$     we have 
\[
 J_{\bm n^\prime}(\sigma)
 =\mathrm{Tr}\!\left[(\hat P_\sigma\otimes\hat I^{\otimes(n-r)})
 \hat P_\nu^\dag\hat\varrho^{(l)}\hat P_\nu\right], \quad \sigma\in S_r,
\]
where $\nu(i)=\alpha_i$ for $1\le i\le r$.

 Eqs.~(\ref{probmar_vis})-(\ref{rvis})    give another, more   physically transparent,  expression  for  the    marginal probability  in Eq.~(\ref{Prn}):
 \be
  p^{(r|n)}_{\bm{m}^{\prime},\bm{n}}=   \binom{n}{r}^{-1} \sum_{\bm{\alpha}} p_{\bm{m}^{\prime},\bm{n}^\prime},
 \en{marprob2}
 where   $p_{\bm{m}^{\prime},\bm{n}^\prime}$ is the $r$-particle probability of   Eq.~(\ref{prob_{nm}})   for the  visible state  in Eq.~(\ref{rvis}).

In Eq.~(\ref{marprob2})     the  marginal  probability is a   uniform  average over all  $r$-particle states, Eq.~(\ref{rvis}), where each   term depends only on the indistinguishability of  $r$   particles.  This observation suggests considering the  corresponding   hierarchy of    indistinguishability. 

\subsection{Hierarchy of    measures of indistinguishability}
Consider  the  projective    measure of   indistinguishability of the  marginal $r$-particle state of Eq.~(\ref{mar_vis}):
\begin{eqnarray}
&&\mathcal{D}^{(r|n)}(\hat{\varrho}) := \mathrm{Tr}\left( \hat{S}^{(\pm)}\hat{\varrho}^{(r|n)}_{(vis)}\right)\nonumber\\
&&= \binom{n}{r}^{-1} \sum_{\bm{\alpha}} \mathrm{Tr} \left( \hat{S}^{(\pm)} \hat{\varrho}^{(\bm{n}^\prime)}_{(vis)}\right)\nonumber\\
&&= \binom{n}{r}^{-1} \sum_{\bm{\alpha}} \frac{1}{r!}\sum_{\sigma\in S_r} J_{\bm{n}^\prime}(\sigma),
\label{Dmar}\end{eqnarray}
where the last expression is obtained by using  Eq.~(\ref{DJ})   applied to the state in Eq. (\ref{rvis}). 

The marginal projective measure $\mathcal{D}^{(r|n)}(\hat{\varrho})$ of Eq.~(\ref{Dmar}) is the complement of the trace distance between the marginal state Eq.~(\ref{mar_vis})  and its normalized ideal projection, provided that the projection has nonzero weight.    The following result  holds (see details in Appendix \ref{appD}). 
\begin{theorem}
 The set of marginal projective measures   Eq.~(\ref{Dmar})  for different $r$   is a   hierarchy  of  upper bounds on the projective measure of indistinguishability of $n$ identical particles:
\be
\mathcal{D}^{(r_2|n)}(\hat{\varrho})\le \mathcal{D}^{(r_1|n)}(\hat{\varrho}), \quad \forall r_1< r_2\le n,
\en{hierD}
where  $\mathcal{D}^{(n|n)}(\hat{\varrho})=\mathcal{D}(\hat{\varrho})$. 
\end{theorem}

A large value of a low-order marginal measure alone need not imply a large $n$-particle measure. For example, the normalized character $J=\chi_{(n-1,1)}/(n-1)$ gives $\mathcal D^{(n|n)}=0$ and $\mathcal D^{(r|n)}=(n-r)/(n-1)$ for $1\le r\le n$. Thus the hierarchy supplies upper bounds, rather than a general approximation guarantee for $\mathcal D^{(n|n)}$. Physically, this conclusion follows from the fact that the $r$-particle marginal probabilities do not  depend on  the higher-order ($\ge r+1$) multi-particle interferences manifested by  the collective multi-particle phases \cite{3phPhase,DistPhInter,nphPhases}. 

 The pairwise bound ($r=2$)  and higher-marginal bounds were also discussed in Ref.~\cite{DistNew}; here the    marginal projective measures  connect the hierarchy to marginal probabilities with possible  experimental readout (discussed below).

\subsection{Marginal projective measures for a stable source of single particles}

As we have discussed in Section \ref{sec4}, independent emissions from a stable source with no inter-emission memory correspond to an $n$-particle label state that is a tensor power of a single-particle label state.  This is a special case of the class-function indistinguishability  with the $n$-particle label state being a convex mixture of tensor-power label  states. For this particular  case we have the following result (see details in Appendix \ref{appD}). 

\begin{proposition}
Let the particles occupy distinct orthogonal visible modes and let their $n$-particle label state be a convex mixture of tensor powers,
\be
    \hat\varrho^{(l)}_n:    =
    \int \hat\rho^{\otimes n}\,d\mu(\hat\rho),
\en{lab_mix_ten}
where $d\mu$ is an arbitrary  probability measure over one-particle  states. Then,  for $n\ge2$,  the corresponding  indistinguishability measure $\mathcal{D}_n$ satisfies the lower bound
 \begin{equation}
    \mathcal D_n
    \ge
    \left(2\mathcal D^{(2|n)}-1\right)^{n-1},  \quad  \mathcal D^{(2|n)}=\mathcal{D}_2. 
    \label{eq:Dn_D2_bound}
\end{equation}
 \end{proposition}
For all $r\le n$ we    have   $\mathcal D^{(r|n)}=\mathcal D_r$ for the label  state in Eq.~(\ref{lab_mix_ten}). 


\subsection{Experimental  readout of   marginal projective measures for bosons}

For $n$ input  bosons in different visible input modes  their  marginal projective measures  can be estimated  directly from an interference experiment  in an  interferometer  of size $M\ge n$  having $n$  balanced inputs for at least one output port, say $l=1$, i.e., for an  interferometer matrix  (see Eqs.~(\ref{Uab2})-(\ref{Uab1}))  satisfying    
\be
 |U_{1,1}| =  |U_{2,1}| = \ldots = |U_{n,1}|=X>0. 
\en{Ubal}
Indeed, consider the marginal boson bunching probability that a uniformly selected subset of $r\le n$ detected bosons lies in output mode $l=1$ ($m^\prime_{1}=r$ and $m^\prime_l = 0$ for $l>1$)  for the  single-particle  input occupation   vector  $\bm{n} = (1,\ldots,1,0,\ldots, 0)$,  given by  Eqs.~(\ref{prob_{nm}}) and (\ref{marprob2}): 
\begin{eqnarray}
\label{BB}
  && p^{(r|n)} (\bm{m}^\prime,\bm{n}) =    \binom{n}{r}^{-1}  \sum_{\bm{n}^\prime\subset \bm{n}} p_{\bm{m}^\prime,\bm{n}^\prime} \nonumber\\
  &&= \binom{n}{r}^{-1} \sum_{\bm{n}^\prime\subset \bm{n}} \frac{1}{r!} \sum_{\sigma,\pi\in S_r} J_{\bm{n}^\prime}(\pi\sigma^{-1}) \prod_{i=1}^rU_{k_{\sigma(i)},l_1}U^*_{k_{\pi(i)},l_1}\nonumber\\
  && = \binom{n}{r}^{-1}  X^{2r} \sum_{\bm{n}^\prime\subset \bm{n}} \sum_{\tau \in S_r} J_{\bm{n}^\prime}(\tau) \nonumber\\
  && = {r!}X^{2r}  \mathcal{D}^{(r|n)}(\hat{\varrho}),
 \end{eqnarray}  
  where  the input occupation vector  $\bm{n}^\prime$ (unique for each  $\bm{\alpha}$ in Eq.~(\ref{marprob2})) runs over all subsets of $r$ particles and  $\tau := \pi\sigma^{-1}$ (we have used that the product of matrix elements with  the same  output port is invariant under permutations).  Therefore, inverting Eq.~(\ref{BB}) by  using the expression of Eq.~(\ref{Prn}) for the marginal probability 
  $p^{(r|n)}_{\bm{m}^\prime,\bm{n}}$, we obtain the  following  expression 
  \be
  \mathcal{D}^{(r|n)}(\hat{\varrho}) = \frac{ p^{(r|n)}_{\bm{m}^\prime,\bm{n}}}{r!X^{2r}} = \frac{(n-r)!}{n!X^{2r}}     \sum_{\bm{m}} \binom{m_1}{r} p_{\bm{m}, \bm{n}},
  \en{RO}
 where   the   summation is over all outputs  with $\bm{m} = (m_1,\ldots, m_M)$ such that  $m_{1}\ge r$.   
  
  Eq.~(\ref{RO}) suggests  the experimental   estimator  $ \widehat{ \mathcal{D}}^{(r|n)}$ of the    marginal projective measure as follows 
 \be
\widehat{ \mathcal{D}}^{(r|n)} = \frac{X^{-2r}}{N}\sum_{i=1}^N F_i, \quad F_i: = \frac{(n-r)!}{n!}\binom{m^{(i)}_1}{r},
 \en{estiDmar}
 where $N$ independent runs are assumed. Since
$0\le F_i\le1/r!$ and $\mathbb \langle F_i\rangle =X^{2r}\mathcal D^{(r|n)}$, this estimator is unbiased and satisfies
\be
 \mathrm{Var}\bigl(\widehat{\mathcal D}^{(r|n)}\bigr)
 \le\frac{X^{-2r}\mathcal D^{(r|n)}}{Nr!}
 \le\frac{X^{-2r}}{Nr!},
 \en{variance_readout}
where we  have used $\mathrm{Var}(F_i) \le \langle F_i\rangle/r!$.  Chebyshev's inequality gives absolute error at most $\eta>0$ with failure probability at most $\delta>0$ whenever
$N\ge X^{-2r}/(r!\eta^2\delta)$.
Thus $N=\mathcal O(X^{-2r})$ is a sufficient bound for fixed $r$, absolute accuracy, and confidence. For a Fourier multiport, $X^2=1/M$ and the estimate is $\mathcal O(M^r)$. 
  
For multiple input mode occupations, the   marginal  boson bunching  probability, generalizing that of Eq.~(\ref{BB}),  reads
\be
p^{(r|n)} (\bm{m}^\prime,\bm{n}) =  \binom{n}{r}^{-1} X^{2r}\sum_{\bm{\alpha}} \frac{1}{\bm{n}^\prime!} \sum_{\tau \in S_r} J_{\bm{n}^\prime}(\tau).
\en{GBB}
Averaging in Eq.~(\ref{GBB}) over the  subsets  of $r$ particles,  as compared to that in Eq.~(\ref{Dmar}), underweights the subsets with multiply occupied modes by the corresponding  multinomial factor  in contrast to the case of a single boson per mode. In this case  the marginal projective measures  would require reconstruction methods for experimental estimation. 
  
We leave as an open problem finding  a  direct fermionic analogue of the bosonic  balanced single-output bunching protocol. General reconstruction of permutation-invariant visible states provides a less direct route to estimating the corresponding projective measures. 

\section{Conclusion}
\label{sec6}
We have  shown that  partial indistinguishability of identical particles with invariant degrees of freedom is defined by a  joint quantum state  of the particles  in the  invariant Hilbert space, which we have termed the label state. The indistinguishability function on the symmetric group encodes the permutation properties of the dynamically invariant label state.  We have given the explicit reconstruction formula for the visible state of $n$ identical particles  in orthogonal input modes  with a fixed occupation vector in terms of  the  indistinguishability function.   

We have  related the projective measures of indistinguishability  and  of emulation of one species by the other  to  the invariant   label state.   Furthermore, we have introduced the complete set of symmetry-sector measures, which completely characterize class-function indistinguishability,  occurring, e.g., for a  permutation-invariant association of labels to particles, such as in experiments involving independent emissions from a stable source of single identical particles.   For fixed distinct occupied modes and class-function indistinguishability, we have characterized maximal distinguishability by the Plancherel distribution of the generalized projective measures. For general indistinguishability functions, the full Fourier matrices are required. We have also presented states which  resolve  a  problem posed in  Ref. \cite{DistNew}.   

We have generalized the indistinguishability theory to coherent superpositions of complete states of identical particles and given a sufficient condition for the corresponding visible part to be a convex mixture. If its components are maximally distinguishable and the cross-indistinguishability functions vanish, the maximal distinguishability is preserved. Our  extension of the Hong--Ou--Mandel criterion for the maximal distinguishability  is distinct from the spectral definition of Ref.~\cite{DistNew}, although both criteria are invariant under the label-independent linear evolution considered here. 

We have introduced a sequence  of marginal projective measures of indistinguishability of identical particles and have shown that it forms a hierarchy of upper  bounds on the projective measure of indistinguishability. Furthermore, we have  provided an operational interpretation of the hierarchy in terms of marginal visible states and marginal probabilities.   Finally,  we have proposed  a   way to obtain upper bounds on the projective measure of indistinguishability for an arbitrarily large number of interfering photons by direct  readout of marginal projective measures of indistinguishability   from an  interference experiment  on a unitary interferometer with at least  one  output port with a uniform single-photon  transition probability. 
 
 \acknowledgments
   This work has been financially supported by  the National Council for Scientific and Technological Development (Conselho Nacional de Desenvolvimento Cient\'ifico e Tecnol\'ogico) of Brazil,  grant number 307507/2023-8.   AI tools have been  used  for proofreading.

\onecolumngrid
\appendix
\section{Proof of Theorem 1 and Corollary 1  of Section \ref{sec2}  }
\label{appA}

To prove Theorem~1, the key step is to  decompose the difference 
$\hat{\varrho}_{(vis)}^{(i)} - \hat{\varrho}_{(vis)}$ into its positive semidefinite and negative semidefinite parts  using  the complementary projectors $\hat{S}^{(\pm)}$ and  $\hat{I}-\hat{S}^{(\pm)}$, using the fact that $[ \hat{S}^{(\pm)} ,\hat{\varrho}_{(vis)}]=0$, by the particle permutation invariance of visible states in Eq.~(\ref{permvisST}) of Section \ref{sec2}, and  $ \hat{S}^{(\pm)} \hat{\varrho}_{(vis)}^{(i)}= \hat{\varrho}_{(vis)}^{(i)}$:
\begin{eqnarray}
\label{Delt_rhoA}
\hat{\varrho}_{(vis)}^{(i)} - \hat{\varrho}_{(vis)}& = & \hat{S}^{(\pm)} \left( \hat{\varrho}_{(vis)}^{(i)} - \hat{\varrho}_{(vis)}\right)+ \left( \hat{I} - \hat{S}^{(\pm)} \right)\left( \hat{\varrho}_{(vis)}^{(i)} - \hat{\varrho}_{(vis)}\right) \nonumber\\
&= &\left( \hat{\varrho}_{(vis)}^{(i)} - \hat{S}^{(\pm)}\hat{\varrho}_{(vis)}\hat{S}^{(\pm)} \right)  
-\left( \hat{\varrho}_{(vis)} - \hat{S}^{(\pm)}\hat{\varrho}_{(vis)}\hat{S}^{(\pm)} \right).
\end{eqnarray}
The first term is $(1-\mathcal D)\hat\varrho^{(i)}_{(vis)}\ge0$, and the second subtracted term is positive semidefinite on the orthogonal complementary sector. Hence, by the definition of the trace distance \cite{BookNC},
\[
d(\hat{\varrho}_{(vis)},\hat{\varrho}_{(vis)}^{(i)}) = \mathrm{Tr}\left( \hat{\varrho}_{(vis)}^{(i)} - \hat{S}^{(\pm)}\hat{\varrho}_{(vis)}\hat{S}^{(\pm)}\right).
\]
Therefore
\begin{eqnarray}
\label{Trd_indA}
  \mathcal{D} (\hat{\varrho}):=1- d(\hat{\varrho}_{(vis)},\hat{\varrho}_{(vis)}^{(i)}) 
  = 1- \mathrm{Tr}\left( \hat{\varrho}_{(vis)}^{(i)} - \hat{S}^{(\pm)}\hat{\varrho}_{(vis)}\hat{S}^{(\pm)}\right)
=   \mathrm{Tr}\left( \hat{S}^{(\pm)}\hat{\varrho}_{(vis)} \right).\nonumber
\end{eqnarray}
The multiplicative property $\vare(\sigma)\vare(\pi) = \vare(\sigma\pi)$ implies the identity \cite{BSF}
\be
\left(\hat{S}^{(\pm)}_{(vis)} \otimes \hat{I}\right) \hat{S}^{(\pm)} =  \left(\hat{I}\otimes \hat{S}^{(+)}_{(int)}\right)\hat{S}^{(\pm)} =\hat{S}^{(\pm)}_{(vis)} \otimes \hat{S}^{(+)}_{(int)}.
\en{id_SA}
Since  $\hat{S}^{(\pm)}\hat{\varrho} = \hat{\varrho}$, Eq.~(\ref{id_SA}) yields 
 \[
 \left(\hat{S}^{(\pm)}_{(vis)} \otimes\hat{I}\right)\hat{\varrho} =\left(\hat{I}\otimes\hat{S}^{(+)}_{(int)} \right)\hat{\varrho}.
 \]
 Taking the partial traces over the visible and internal subspaces  gives 
\be
\mathcal{D} (\hat{\varrho}) = \mathrm{Tr}\left( \hat{S}^{(\pm)}\hat{\varrho}_{(vis)} \right) =\mathrm{Tr}\left( \hat{S}^{(+)}\hat{\varrho}_{(int)}\right),
\en{DintA}
which is   Theorem~1.  To prove Corollary~1 we  employ $\hat S^{(\mp)}$  and the analogous identity  $(\hat S^{(\mp)}_{(vis)}\otimes\hat I)\hat S^{(\pm)}=(\hat I\otimes\hat S^{(-)}_{(int)})\hat S^{(\pm)}$.  Q.E.D.

\section{ Proof of Theorem~2 and Proposition 1  of Section \ref{sec3}}
\label{appB}
\subsection{Proof of the direct result in Theorem~2 for  the simplest case}

We first consider the simpler case of $n$ identical particles, originating from independent sources, and  occupying mutually orthogonal visible modes $|k\rangle\in \mathcal{H}_{(vis)}$, $k=1,\ldots, n$.    Their label states are denoted by $\hat{\varrho}^{(l)}_k$ (in quantum optics, the label state is usually referred to simply as the ``state'' of a photon). Using a basis $|j\rangle$, $j=1,2,3,\ldots$, of the internal Hilbert space $\mathcal{H}_{(int)}$ we   expand  
\be
\hat{\varrho}^{(l)}_k = \sum_{j,l}\varrho_{k;jl}|j\rangle\langle l|.
\en{ED1} 
With creation operators  $\hat{a}^\dag_{k,j}$ and annihilation operators  $\hat{a}_{k,j}$  for  the single-particle state  $|k\rangle\otimes |j\rangle$, 
the many-particle state in the second-quantization representation becomes
\be
\hat{\varrho} = \sum_{j_1,\ldots,j_n}\sum_{l_1,\ldots,l_n} \left( \prod_{k=1}^n\varrho_{k;j_k,l_k}\right) \prod_{k=1}^n \hat{a}^\dag_{k,j_k}|0\rangle\langle 0| 
\left(\prod_{k=1}^n \hat{a}^\dag_{k,l_k}\right)^\dag.
\en{ED2}
Conversely, any state of the form in  Eq.~(\ref{ED2}) uniquely determines the particle-label states $\hat{\varrho}^{(l)}_k$.
Using the relation between the second- and first-quantization representations,
\be
  \prod_{k=1}^n\hat{a}^\dag_{\phi_k}|0\rangle = \sqrt{n!} \hat{S}^{(\pm)} \bigotimes_{k=1}^n |\phi_k\rangle,
\en{ED3}
valid for an arbitrary set of single-particle states  $|\phi_k\rangle\in \mathcal{H}_{(vis)}\otimes \mathcal{H}_{(int)}$  (for a   proof, see, e.g., Ref. \cite{LecNotes}),  Eq.~(\ref{ED2})  becomes 
\be
\hat{\varrho}  = n! \sum_{j_1,\ldots,j_n}\sum_{l_1,\ldots,l_n} \left( \prod_{k=1}^n\varrho_{k;j_k,l_k}\right) \hat{S}^{(\pm)} \left\{\bigotimes_{k=1}^n |k\rangle\langle k| \otimes |j_k\rangle\langle l_k|\right\} \hat{S}^{(\pm)} =n! \hat{S}^{(\pm)} \left\{\bigotimes_{k=1}^n |k\rangle\langle k|\otimes \hat{\varrho}^{(l)}_k\right\} \hat{S}^{(\pm)}.
\en{ED4}
Hence, in the first-quantization representation, the corresponding   seed state of non-identical particles and the associated label state are
 \be
\hat{\varrho}^{(d)} \equiv  \bigotimes_{k=1}^n |k\rangle\langle k|\otimes      \hat{\varrho}^{(l)}_k= \left( \bigotimes_{k=1}^n |k\rangle\langle k|\right)\otimes \hat{\varrho}^{(l)}, \quad \hat{\varrho}^{(l)}\equiv \bigotimes_{k=1}^n\hat{\varrho}^{(l)}_k.
\en{varrho^d}
	
We now derive the corresponding visible state. Tracing over  the internal degrees of freedom in  Eq.~(\ref{ED4}),  expanding the projectors $\hat{S}^{(\pm)}$,  and using the group identities  $\hat{P}^\dag_\pi = \hat{P}_{\pi^{-1}}$ and $\hat{P}_{\pi^{-1}}\hat{P}_\sigma = \hat{P}_{\pi^{-1}\sigma}$, we obtain 
\begin{eqnarray}
\hat{\varrho}_{(vis)}=\mathrm{Tr}_{(int)}\hat{\varrho} &= & \frac{1}{n!} \sum_{\sigma,\pi\in S_n} \vare(\pi^{-1}\sigma) \mathrm{Tr}
\left(\hat{P}^\dag_\pi\hat{P}_\sigma   \hat{\varrho}^{(l)}  \right) \hat{P}_\sigma  \left(\bigotimes_{k=1}^n |k\rangle\langle k| \right) \hat{P}^\dag_\pi\nonumber\\
&=& \frac{1}{n!} \sum_{\sigma,\pi\in S_n} \vare(\pi\sigma)  J_{\hat{\varrho}}(\pi^{-1}\sigma)  \bigotimes_{k=1}^n |\sigma^{-1}(k)\rangle\langle \pi^{-1}(k)|\nonumber\\
&=& \frac{1}{n!} \sum_{\sigma,\pi\in S_n} \vare(\pi\sigma)  J_{\hat{\varrho}}(\pi \sigma^{-1})  \bigotimes_{k=1}^n |\sigma(k)\rangle\langle \pi (k)|,
\label{ED5}
\end{eqnarray}
where the indistinguishability function is
\be
J_{\hat{\varrho}}(\sigma) :=    \mathrm{Tr}\left(\hat{P}_\sigma \hat{\varrho}^{(l)} \right)=\mathrm{Tr}\left(\hat{P}_\sigma \bigotimes_{k=1}^n\hat{\varrho}^{(l)}_k \right) = \prod_{\nu\in cyc(\sigma)} \mathrm{Tr}\left(   \hat{\varrho}^{(l)}_{k_1}\cdot \ldots \cdot   \hat{\varrho}^{(l)}_{k_{|\nu|}}\right),
\en{ED6}
where $\nu$ is a cycle of length  $|\nu|$ in the disjoint  cycle decomposition of $\sigma$, acting as follows   $\nu(k_i) = k_{i-1}$, $ i=2,\ldots, |\nu|$ and $\nu(k_1) = k_{|\nu|}$.  
  Finally, taking the partial trace of Eq.~(\ref{ED4}) over the visible degrees of freedom and using the orthogonality of the visible modes yields the relation between the label state and the internal state,
\be
\hat{\varrho}_{(int)}=\mathrm{Tr}_{(vis)}\hat{\varrho} =\frac{1}{n!}\sum_{\sigma\in S_n} \hat{P}_\sigma \hat{\varrho}^{(l)}\hat{P}^\dag_\sigma.
\en{REL1}


\subsection{Proof of the direct result in Theorem~2 for the general case}
 
Utilizing Eq. (\ref{ED3}) in a similar way, we obtain the visible state corresponding to $|\Psi_{\bm{n}}\rangle$ in Eq. (\ref{Psi_nSQ}) of the main text:
\begin{eqnarray}
\label{ED10}
\hat{\varrho}_{(vis)} &= &\mathrm{Tr}_{(int)}\{ |\Psi_{\bm{n}}\rangle\langle \Psi_{\bm{n}}|\} = \frac{1}{n!\bm{n}!} \sum_{\sigma,\pi\in S_n} \vare(\pi\sigma)  J_{\hat{\varrho}}(\pi^{-1}\sigma)  \bigotimes_{\alpha=1}^n |k_{\sigma^{-1}(\alpha)}\rangle\langle k_{\pi^{-1}(\alpha)}|\nonumber\\
&=&\frac{1}{n!\bm{n}!} \sum_{\sigma,\pi\in S_n} \vare(\pi\sigma)  J_{\hat{\varrho}}(\pi\sigma^{-1})  \bigotimes_{\alpha=1}^n |k_{\sigma(\alpha)}\rangle\langle k_{\pi(\alpha)}|,
\end{eqnarray} 
where the indistinguishability function of the state  $|\Psi_{\bm{n}}\rangle$ is 
\be
 J_{\hat{\varrho}}(\sigma) := \left[ \sum_{l_1,\ldots,l_n}C^*_{l_1,\ldots,l_n} \bigotimes_{\alpha=1}^n \langle l_\alpha| \right]   \hat{P}_\sigma \left[ \sum_{j_1,\ldots,j_n}C_{j_1,\ldots,j_n}\bigotimes_{\alpha=1}^n |j_\alpha\rangle\right] = \mathrm{Tr}\{ \hat{P}_\sigma \hat{\varrho}^{(l)}\},
\en{ED11}
with the corresponding label state 
\be
\hat{\varrho}^{(l)} := |\Psi^{(l)}_{\bm{n}}\rangle \langle \Psi^{(l)}_{\bm{n}}|, \quad |\Psi^{(l)}_{\bm{n}}\rangle:= \sum_{j_1,\ldots,j_n}C_{j_1,\ldots,j_n}\bigotimes_{\alpha=1}^n |j_\alpha\rangle.
\en{ED12}
 
 The  symmetry with respect to the Young subgroup $\mathcal{Y}_{\bm{n}}$,  induced by the occupation vector $\bm{n}$, Eq.~(\ref{ED8}) of the main text, is inherited by the label state and, consequently, by the indistinguishability function. In particular, 
\be
\hat{P}_\pi |\Psi^{(l)}_{\bm{n}}\rangle =  \vare(\pi) |\Psi^{(l)}_{\bm{n}}\rangle, \quad \forall \pi \in \mathcal{Y}_{\bm{n}}.
\en{ED13}
Taking the partial trace over the visible degrees of freedom of the complete state obtained from Eqs.~(\ref{Psi_nSQ}) of Section \ref{sec3A} and (\ref{ED3}), and using the Young-subgroup symmetry, gives Eq.~(\ref{REL1}).  Q.E.D.

The label state in Eq.~(\ref{ED12}) can account for arbitrary entanglement of the particles over the internal degrees of freedom and, in general, is therefore not factorized, unlike the simpler case of one particle per visible mode considered above. Multiple occupation imposes the Young-subgroup symmetry in Eq.~(\ref{ED13}), but does not by itself force entanglement for bosons: a tensor power of one pure label state is already symmetric. For fermions, the required antisymmetry within a multiply occupied mode excludes a pure product of the corresponding one-particle label states.

\subsection{Proof of the converse result in Theorem 2}
Consider an arbitrary visible state $\hat{\varrho}_{(vis)}$ of $n$ identical particles with the  occupation vector $\bm{n}=(n_1,\ldots,n_M)$, with $n_{r+1}=\cdots=n_M=0$, over mutually orthogonal visible modes $|1\rangle, \ldots |r\rangle$. 
Thanks to the    particle permutation symmetry  of the visible state, Eq.~(\ref{permvisST}) of Section \ref{sec2}, it  admits the following decomposition
\be
\hat{\varrho}_{(vis)} = \frac{1}{n!\bm{n}!}\sum_{\sigma,\pi\in S_n} M_{\sigma,\pi} \hat{P}_\sigma \Biggl(\bigotimes\limits_{k=1}^r \left(|k\rangle\langle k|\right)^{\otimes n_k}\Biggr)\hat{P}^\dag_\pi, 
\en{formSt}
where the matrix $M$ indexed by permutations is positive semidefinite and Hermitian. An explicit choice is
\[
 M_{\sigma,\pi}=\frac{n!}{\bm n!}
 \langle K|\hat P_\sigma^\dag\hat\varrho_{(vis)}\hat P_\pi|K\rangle,
 \qquad |K\rangle:=\bigotimes_{k=1}^r|k\rangle^{\otimes n_k}.
\]
Each distinct occupation-basis vector occurs $\bm n!$ times in the permutation sum, which verifies the normalization factor in Eq.~(\ref{formSt}).  Since  
\be
\hat{P}_\tau \bigotimes\limits_{k=1}^r |k\rangle^{\otimes n_k} =  \bigotimes\limits_{k=1}^r |k\rangle^{\otimes n_k} , \quad \forall \tau \in \mathcal{Y}_{\bm{n}},
\en{symV}
 where $\mathcal{Y}_{\bm{n}}=S_{n_1}\times\ldots \times S_{n_r}$ is the Young subgroup induced by the occupation vector  $\bm{n}$, without changing the state, one may choose the coefficients to satisfy 
  \be
 M_{\sigma\tau,\pi} = M_{\sigma,\pi\tau} = M_{\sigma, \pi}, \quad \quad \forall \tau \in \mathcal{Y}_{\bm{n}}. 
 \en{symM}
 Using Eq.~(\ref{symM}), the normalization condition becomes
  \be
\frac{1}{n! \bm{n}!} \sum_{\sigma\in S_n}\sum_{\tau\in \mathcal{Y}_{\bm{n}}} M_{\sigma,\sigma\tau}=1. 
 \en{normM}
For this choice of $M$, Eq.~(\ref{permvisST}) gives 
 \be
 M_{\mu\sigma,\mu\pi} = M_{\sigma,\pi},  \quad \forall \mu \in S_n,
 \en{symM2}
setting  $\mu = \pi^{-1}$  we get  $M_{\sigma,\pi} = M_{\pi^{-1}\sigma,e}$, while  for $\mu =\sigma^{-1}$ we get $M_{\sigma,\pi} =M_{e,\sigma^{-1}\pi}$. Now we set 
\be
\Lambda(\sigma) := M_{\sigma,e} = M_{e,\sigma^{-1}} .
\en{JfromM}
By definition, the  function in  Eq.~(\ref{JfromM}) is positive semidefinite  on $S_n$, symmetric under the Young subgroup,   
\be
 \Lambda(\sigma\tau)= \Lambda(\tau \sigma) = \Lambda(\sigma),\quad    \forall \tau \in \mathcal{Y}_{\bm{n}},
\en{symJM}
and  normalized by $\Lambda(e)=1$.   Introducing the indistinguishability function   $J(\sigma) := \Lambda(\sigma)$ in the case of  bosons and $J(\sigma) := \mathrm{sgn}(\sigma)\Lambda(\sigma)$ in the case of fermions and the  non-decreasing mode sequence $k_\alpha \le k_{\alpha +1}$, corresponding  to the occupation vector $\bm{n}$, we  arrive at the form of the state in Theorem 2.  Q.E.D. 

The converse result in  Theorem~2  is further  strengthened by  Proposition 1 of Section \ref{sec3A}, which is proven as follows.

\subsection{Proof of Proposition 1  }
 Let us explicitly construct   a  quantum  state of  Proposition 1 from     $J(\sigma)$. Consider a Hilbert space $\mathcal{H}$ of dimension $\mathrm{dim}(\mathcal{H}) \ge n$ and choose some  orthogonal states  $|j\rangle\in \mathcal{H}$, $j=1,\ldots, n$. Introduce the following set of $n!$ orthogonal states 
\be
|\sigma\rangle:= \hat{P}_\sigma\bigotimes_{j=1}^n |j\rangle = \bigotimes_{j=1}^n |\sigma^{-1}(j)\rangle. 
\en{sigmaST}
Now,  we can  use  in Eq. (\ref{Jform}) of Section \ref{sec3A}  the quantum   state  
\be
\hat{\varrho}^{(J)} = \frac{1}{n!}\sum_{\tau, \pi\in S_n} |\pi\rangle J(\tau\pi^{-1})\langle \tau|.
\en{mixStJ}
Indeed, the  operator $\hat{\varrho}^{(J)}$ is positive semidefinite  because $J(\sigma)$ is positive semidefinite; it also  has unit trace, because the states $|\sigma\rangle$ are orthonormal and $J(e)=1$.    Using that
\[
\hat{P}_\sigma |\pi\rangle = \bigotimes_{j=1}^n |\pi^{-1}(\sigma^{-1}(j))\rangle =\bigotimes_{j=1}^n |(\sigma\pi)^{-1}(j)\rangle= \hat{P}_{\sigma\pi}|e\rangle = |\sigma\pi\rangle,
\]
we obtain
\[
\mathrm{Tr}(\hat{P}_\sigma \hat{\varrho}^{(J)}) = \frac{1}{n!}\sum_{\tau, \pi\in S_n}   J(\tau\pi^{-1})\delta_{\tau,\sigma\pi} = J(\sigma).
\]
Q.E.D.

\section{Proof of Proposition 2 and Theorem 3 of Section \ref{sec4}  }
\label{appC}

 Class functions have the same values on all permutations of the same cycle type $\lambda$, e.g., $\chi_\lambda(\sigma) = \chi_\lambda(\sigma^{-1})$. The generalized  orthogonality property of the irreducible characters follows from that of the irreducible representations and  is as follows (we also take into account that the irreducible characters are real-valued):
\be
\frac{1}{n!}\sum_\sigma \chi_\lambda(\sigma)\chi_\mu(\tau\sigma) = \frac{\delta_{\lambda,\mu}\chi_\lambda(\tau)}{\chi_\lambda(e)},
\en{chi_orth}
where $\lambda$ and $\mu$ are two partitions of $n$. Setting $\tau = e$ we get the mutual orthogonality property with respect to the inner product of two complex-valued class functions, $f(\sigma)$ and $g(\sigma)$ on $S_n$, defined as follows
\be
(f,g):= \frac{1}{n!}\sum_{\sigma\in S_n} f^*(\sigma)g(\sigma). 
\en{Inn_prod}
The orthogonality property  Eq. (\ref{chi_orth}) allows us to  introduce projectors $\hat{S}^{(\lambda)}$,  given by Eq. (\ref{S_lambda}) of Section \ref{sec4},
acting on the $n$th  tensor power of a Hilbert space and satisfying 
\be
\quad \hat{S}^{(\lambda)} \hat{S}^{(\mu)}= \delta_{\lambda,\mu}\hat{S}^{(\lambda)}, \quad \sum_{\lambda\vdash n}\hat{S}^{(\lambda)} =\hat{I}.
\en{SlambSmu}
The associated irreducible representations      are multidimensional, with the exceptions being  the one-dimensional trivial (indistinguishable bosons) and sign (indistinguishable fermions)  representations.

\subsection{Proof of Proposition 2}
Since the  irreducible characters of $S_n$ are a basis for the class functions on $S_n$, any indistinguishability function $J(\sigma)$ that is a class function is expandable as 
\be
J(\sigma) = \sum_{\lambda\vdash n} c_\lambda \frac{\chi_\lambda(\sigma)}{\chi_\lambda(e)},
\en{expchar}
with some $c_\lambda$. By the orthogonality of the characters with respect to the inner product in Eq.~(\ref{Inn_prod})  we have the following expression 
\be
c_\lambda = \chi_\lambda(e)(\chi_\lambda, J). 
\en{clambda}
Now, to show that $c_\lambda \ge 0$, we use Proposition 1 of Section \ref{sec3A} to cast our  indistinguishability function $J(\sigma)$ using the associated  quantum state $\hat{\varrho}^{(J)}= \sum_{j} q_j |\Phi_j\rangle \langle \Phi_j|$ (expanded over the  eigenstates)
\be
J(\sigma) = \mathrm{Tr}\left(\hat{P}_\sigma \hat{\varrho}^{(J)} \right) = \sum_{j} q_j \langle \Phi_j|\hat{P}_\sigma|\Phi_j\rangle.
\en{J4rho}
With the expansion of Eq.~(\ref{J4rho}) substituted into the expression of Eq.~(\ref{clambda}) and with the use of the orthogonal projectors Eq.~(\ref{S_lambda}) on the symmetry sectors, we get
\begin{eqnarray}
 c_\lambda = \chi_\lambda(e)\sum_j q_j \langle \Phi_j| \frac{1}{n!}\sum_{\sigma\in S_n} \chi_\lambda (\sigma) \hat{P}_\sigma|\Phi_j\rangle= 
\sum_j q_j \langle \Phi_j| \hat{S}^{(\lambda)}|\Phi_j\rangle\ge 0.
 \end{eqnarray}
Evaluating Eq.~(\ref{expchar}) at $e$ gives $\sum_\lambda c_\lambda=J(e)=1$.
 Q.E.D.  

\subsection{Proof of the first part of Theorem 3:  projective measures on symmetry sectors }

The following identity generalizes Eq.~(\ref{id_SA}) from the bosonic and fermionic sectors to arbitrary irreducible symmetry sectors:
\be
 \left(\hat{I}\otimes \hat{S}^{(\lambda)}_{(int)}\right)\hat{S}^{(\pm)} = \left(\hat{S}^{(\lambda^\prime)}_{(vis)} \otimes \hat{I}\right) \hat{S}^{(\pm)}  ,\quad\lambda^\prime 
 =\left\{ \begin{array}{cc} \lambda, & \mathrm{bosons},\\  \lambda^T, & \mathrm{fermions},\end{array} \right.
\en{id_SpmSlambda}
where $\lambda^T$ is the transposed partition (i.e., the transposed Young diagram). Indeed, using the multiplicativity $\vare(\sigma)\vare(\pi) = \vare(\sigma\pi)$, together with the identities  $\mathrm{sgn}(\sigma)\chi_\lambda(\sigma) = \chi_{\lambda^T}(\sigma)$  and    $\chi_{\lambda^T}(e) = \chi_\lambda(e)$ one finds
\be
\left(\hat{I}\otimes \chi_\lambda(\sigma) \hat{P}_\sigma\right)\vare(\pi)\hat{P}_\pi =\left( \vare(\sigma)\chi_\lambda(\sigma) \hat{P}_{\sigma^{-1}} \otimes  \hat{I}\right) \vare(\sigma\pi)\hat{P}_{\sigma\pi} = \left(  \chi_{\lambda^\prime}(\sigma^{-1}) \hat{P}_{\sigma^{-1}}  \otimes\hat{I}\right) \vare(\sigma\pi)\hat{P}_{\sigma\pi}.
\en{id_PP}
Averaging both sides in  Eq. (\ref{id_PP})  over $\sigma$ and $\pi$ we get Eq. (\ref{id_SpmSlambda}).    

Since  $\hat{S}^{(\pm)}\hat{\varrho} = \hat{\varrho}$, Eq.~(\ref{id_SpmSlambda})  implies 
 \[
 \left(\hat{I}\otimes\hat{S}^{(\lambda)}_{(int)} \right)\hat{\varrho} = \left(\hat{S}^{(\lambda^\prime)}_{(vis)} \otimes\hat{I}\right)\hat{\varrho}.
 \]
  Using this relation and taking the partial traces over the visible and internal degrees of freedom, we obtain the result stated in Theorem 3 of Section \ref{sec4}  for an arbitrary partition $\lambda$:
\be
\mathcal{D}^{(\lambda)} (\hat{\varrho}) := \mathrm{Tr}\left( \hat{S}^{(\lambda)}\hat{\varrho}_{(int)}\right)=\mathrm{Tr}\left( \hat{S}^{(\lambda^\prime)}\hat{\varrho}_{(vis)} \right).
\en{D_lam_intA}
Using the fact that the internal state coincides with the  averaged  label state,   
\[
\hat{\varrho}_{(int)} = \frac{1}{n!}\sum_{\tau\in S_n} \hat{P}_\tau  \hat{\varrho}^{(l)} \hat{P}^\dag_\tau,
\]
and that $\chi_\lambda(\sigma)$ is a class function,  we can rewrite  $\mathcal{D}^{(\lambda)} (\hat{\varrho}) $ of  Eq.~(\ref{D_lam_intA})  as follows 
\be
\mathcal{D}^{(\lambda)} (\hat{\varrho})=    \frac{\chi_\lambda(e)}{n!}\sum_{\sigma\in S_n}  \chi_\lambda(\sigma)\mathrm{Tr}\left(\hat{P}_\sigma\frac{1}{n!}\sum_{\tau\in S_n} \hat{P}_\tau  \hat{\varrho}^{(l)} \hat{P}^\dag_\tau \right)
= \frac{\chi_\lambda(e)}{n!}\sum_{\sigma \in S_n} \chi_\lambda(\sigma)J_{\hat{\varrho}}(\sigma) .
\en{D_lambdaJ}
 
The following proposition  summarizes the above results.
 \begin{customprop}
 For every partition $\lambda\vdash n$,  let   the projector onto the isotypic symmetry sector labeled by $\lambda$ be   as follows
\[
\hat {S}^{(\lambda)} = \frac{\chi_\lambda(e)}{n!} \sum_{\sigma\in S_n} \chi_\lambda(\sigma)\hat P_\sigma.
\]
 Then the corresponding projective measure is
\[
\mathcal D^{(\lambda)}(\hat\varrho) := \mathrm{Tr}\!\left(\hat S^{(\lambda)}\hat\varrho_{(int)}\right)
= \mathrm{Tr}\!\left( \hat S^{(\lambda^\prime)}\hat\varrho_{(vis)} \right),
\]
where
$\lambda^\prime=\lambda$ for bosons and $\lambda^\prime=\lambda^T$ for fermions.   The  probability distribution $\{\mathcal D^{(\lambda)},\lambda\vdash n\}$ satisfies  
\[
\mathcal D^{(\lambda)} = \frac{\chi_\lambda(e)}{n!} \sum_{\sigma\in S_n} \chi_\lambda(\sigma)J(\sigma), 
\]
 and provides a complete characterization of the class-function indistinguishability, i.e., when 
 \[
J(\tau^{-1}\sigma\tau)=J(\sigma), \quad \forall\,\sigma,\tau\in S_n.
\]
 \end{customprop}

 \subsection{ Proof of the second part of Theorem 3:   generalized symmetry sectors}
 
Due to the particle permutation symmetry of the visible state, Eq.~(\ref{permvisST}) of Section \ref{sec2}, the visible state is the weighted direct sum of the normalized projected states, where for every partition $\lambda\vdash n$ with $\mathcal D^{(\lambda^\prime)}>0$ as defined in Eq.~(\ref{D_lam_intA}), we have 
 \be
 \hat{\varrho}^{(\lambda)} _{(vis)}:= \frac{1}{\mathcal{D}^{(\lambda^\prime)} }\hat{S}^{(\lambda)} \hat{\varrho}_{(vis)}\hat{S}^{(\lambda)}
 \en{projstlam}
 and sectors of zero weight are omitted. 
Since every $\hat S^{(\lambda)}$ is a central element of the represented group algebra of $S_n$, it commutes with all permutation operators. Therefore,
 \[
 \hat{P}_\sigma  \hat{\varrho}^{(\lambda)}_{(vis)} \hat{P}_\sigma^\dag =  \hat{\varrho}^{(\lambda)}_{(vis)}, \quad  \forall \sigma \in S_n,
 \]
   so each projected state is again a valid visible state of identical particles. Furthermore, the projectors onto distinct irreducible symmetry sectors are mutually orthogonal,
 \[
 \hat{S}^{(\lambda)} \hat{S}^{(\mu)}=\delta_{\lambda,\mu} \hat{S}^{(\lambda)} ,
 \]
 which implies
 \[
  \hat{S}^{(\lambda)} \hat{\varrho}_{(vis)}  \hat{S}^{(\mu)} = 0, \quad  \lambda\ne \mu. 
 \]
Thus the visible state decomposes uniquely into orthogonal generalized symmetry sectors
\be
 \hat{\varrho} _{(vis)} =\bigoplus_{\lambda\vdash n}  \mathcal{D}^{(\lambda^\prime)} \hat{\varrho}^{(\lambda)} _{(vis)}, 
\en{vis_st_lambda1}
where the   probability $\mathcal D^{(\lambda^\prime)}$ gives the weight of the orthogonal component of the  visible state with definite generalized symmetry and  the   state $\hat{\varrho}^{(\lambda)}_{(vis)}$ describes its structure within the corresponding isotypic symmetry sector. This is the symmetry-sector part of the Schur--Weyl decomposition of the visible state.

\subsection{States of $n=3$ particles with  given ($\mathcal{D}, \widetilde{\mathcal{D}}$) }

Consider $n=3$ particles occupying three distinct orthogonal visible modes $|k\rangle$, $k=1,2,3$ and having class-function indistinguishability $J(\sigma)$. Their visible state can be written as
\be
\hat{\varrho}_{(vis)} = \frac{1}{6} \sum_{\sigma,\pi\in S_3}\vare(\pi\sigma)J(\pi^{-1}\sigma)|\sigma\rangle\langle \pi|, 
\en{CS1}
with  $|\sigma\rangle$ given by Eq.~(\ref{sigmaST}) for $n=3$. The symmetric group $S_3$ has three characters, two one-dimensional $\chi_{(3)}(\sigma)=1$ and  $\chi_{(1,1,1)}(\sigma) =  \mathrm{sgn}(\sigma)$ and one two-dimensional $\chi_{(2,1)}(\sigma)$.  The characters  have the same values on permutations of the same conjugacy class, i.e.,  of the same  cycle type. The group $S_3$ has three conjugacy classes, corresponding to the three partitions: the identity $e$, the transpositions $t=\{(1,2), (2,3), (1,3)\}$, and the two $3$-cycles $c=\{(1,2,3),(1,3,2)\}$.   The values of $\chi_{(2,1)}$ are as follows
\be
\chi_{(2,1)}(e) = 2, \quad \chi_{(2,1)}(t) = 0, \quad \chi_{(2,1)}(c) = -1. 
\en{CS2}
Let us write down the explicit matrix form of the state in Eq.~(\ref{CS1}) in the basis $|\sigma\rangle$ ordered by permutation $\sigma$: 
\[
e,\; (1,2), \;(2,3), \; (1,3), \;(1,2,3),\; (1,3,2)
\]
 (i.e., $e$, three transpositions, two $3$-cycles), or in the explicit form $|i_1,i_2,i_3\rangle$ ($i_k:= \sigma^{-1}(k)$) as follows 
\be
|1,2,3\rangle, \; |2,1,3\rangle, \;  |1,3,2\rangle, \;  |3,2,1\rangle, \;  |3,1,2\rangle, \;  |2,3,1\rangle. 
\en{CS3}
The indistinguishability class function in Eq. (\ref{Jnew}) of Section \ref{sec4}   reads  
\be
J(\sigma)= \mathcal{D}  + \widetilde{\mathcal{D}}\mathrm{sgn}(\sigma) + (1- \mathcal{D}-\widetilde{\mathcal{D}})\frac{1}{2}\chi_{(2,1)}(\sigma).
\en{CS4}
From Eq. (\ref{CS2}) we obtain
\be
J(e) = 1, \quad J(t) = \mathcal{D}-\widetilde{\mathcal{D}}, \quad J(c) = \frac{3(\mathcal{D}+\widetilde{\mathcal{D}})-1}{2}. 
\en{CS5}

Since $J(\sigma)$ depends only on the conjugacy class of $\sigma$, the matrix elements are determined entirely by the two parameters  $a = \mathcal{D}-\widetilde{\mathcal{D}}$ and $b = \frac{3(\mathcal{D}+\widetilde{\mathcal{D}})-1}{2}$. Consider, for instance,  bosons $\vare(\sigma)=1$. We obtain the matrix 
$\varrho_{(vis)}$ of the visible state $\hat{\varrho}_{(vis)}$  Eq. (\ref{CS1}) in the   ordered basis of Eq. (\ref{CS3}) as follows
\be
\varrho^{(B)}_{(vis)} = \frac{1}{6}
\begin{pmatrix}
1 & a & a & a & b & b \\
a & 1 & b & b & a & a \\
a & b & 1 & b & a & a \\
a & b & b & 1 & a & a \\
b & a & a & a & 1 & b \\
b & a & a & a & b & 1
\end{pmatrix}.
\en{CS6}
For fermions, due to $\mathrm{sgn}(e,t,c) = (1,-1,1)$,  the respective matrix is 
\be
\varrho^{(F)}_{(vis)} =\mathcal{S} \varrho^{(B)}_{(vis)} \mathcal{S}, \quad \mathcal{S} = \mathrm{diag}(1,-1,-1,-1,1,1).  
\en{fermionCS6}
 
 The matrices in Eqs. (\ref{CS6})-(\ref{fermionCS6}) are  real and obviously symmetric (i.e., Hermitian). To check the  positive semidefiniteness, let us diagonalize $\varrho^{(B)}_{(vis)}$.   The eigenvalues with their multiplicities read:
\be
\lambda_1=\frac{1+3a+2b}{6}=\mathcal{D}, \quad 
 \lambda_2=\frac{1-3a+2b}{6} = \widetilde{\mathcal{D}}, \quad 
 \lambda_3=\frac{1-b}{6} =\frac{1-\mathcal{D}-\widetilde{\mathcal{D}}}{4}\; (\text{multiplicity } 4)
 \en{CS7}
 Therefore, the matrix $\varrho_{(vis)} $ Eq. (\ref{CS6}) is positive semidefinite iff 
 \[
  \mathcal{D}\ge 0,\quad  \widetilde{\mathcal{D}}\ge 0 ,  \quad \mathcal{D}+\widetilde{\mathcal{D}}\le 1.
\]
Since $\mathcal{D},\widetilde{\mathcal{D}}\ge0$ by definition, the remaining condition is $\mathcal{D}+\widetilde{\mathcal{D} }\le1$. Thus for every pair $(\mathcal{D},\widetilde{\mathcal{D}})$  satisfying $\mathcal{D}+\widetilde{\mathcal{D} }\le1$ there is a visible state given by Eqs.~(\ref{CS1}) and (\ref{CS5}), whereas the respective  label state can be the  state of  Eq.~(\ref{mixStJ}). In the matrix form, the latter  coincides   with  the matrix of the visible state of bosons, whereas  for fermions  it is   the $\mathrm{sgn}$-reflection of the respective matrix.

Setting $\mathcal{D}=\frac{1}{6}$ and $\widetilde{\mathcal{D}}\ne \frac{1}{6}$ gives $a\ne0$ and $b\ne0$, so the visible state is manifestly non-diagonal despite having the same   indistinguishability measure as maximally distinguishable particles. It lies outside the convex hull of perfectly distinguishable states  utilized in  Ref.~\cite{DistNew}.
Introduce  the space 
\[
 \mathcal V=\operatorname{span}\{|\sigma\rangle:\sigma\in S_3\},
\]
which  carries one copy of the regular representation of $S_3$ and has dimension six. Let $\mathfrak D$ denote the set of perfectly distinguishable states in the spectral sense of Ref.~\cite{DistNew}. In the remainder of this convex-hull argument, distinguishability refers specifically to that spectral definition. Suppose, for a contradiction, that  the state in  Eq.~(\ref{CS6}) can be expanded as follows 
\[
\varrho^{(B)}_{(vis)}=\sum_i p_i\hat\omega_i,
\qquad p_i>0,\qquad \hat\omega_i\in\mathfrak D,
\]
where the states $\hat\omega_i$ may be diagonal in different bases related by some permutation-invariant unitaries.
Consider $|v\rangle$ such that  $\varrho^{(B)}_{(vis)}|v\rangle=0$, i.e., 
\[
0=\langle v|\varrho^{(B)}_{(vis)}|v\rangle
  =\sum_i p_i\langle v|\hat\omega_i|v\rangle .
\]
Since every term is nonnegative, we get  $\hat\omega_i|v\rangle=0$. Therefore
\[
\operatorname{supp}\hat\omega_i
\subseteq\operatorname{supp}\varrho^{(B)}_{(vis)}
\subseteq\mathcal V
\qquad\text{for every }i.
\]
In Ref.~\cite{DistNew}, due to Schur-Weyl duality,   every nonzero eigenspace of a maximally distinguishable state carries an integer number of regular representations. Hence its dimension is a positive multiple of \(|S_3|=6\). Since every \(\hat\omega_i\) is supported in the six-dimensional space \(\mathcal V\), it can possess only one nonzero eigenspace, which must coincide with \(\mathcal V\). Thus \(\hat\omega_i\) is proportional to the identity on \(\mathcal V\), and normalization gives
\[
\hat\omega_i=\frac{\hat I_{\mathcal V}}6 .
\]
This conclusion is unaffected by allowing different permutation-invariant rotations: any rotated regular subspace contained in \(\mathcal V\) has dimension six and must therefore equal \(\mathcal V\).

It follows that
\[
\sum_i p_i\hat\omega_i=\frac{\hat I_{\mathcal V}}6.
\]
However, for $\mathcal D=1/6$, the eigenvalues of Eq.~(\ref{CS7}) are
\[
\frac16,\qquad
\widetilde{\mathcal D},\qquad
\frac{5/6-\widetilde{\mathcal D}}4
\quad(\text{multiplicity }4).
\]
 Therefore, whenever
\[
\widetilde{\mathcal D}\neq\frac16,
\]
the state in Eq.~(\ref{CS6})  cannot be represented as a convex combination of maximally distinguishable states, even when those states are diagonal in different permutation-invariantly rotated bases. 

Finally, though the  proof  has been carried out  explicitly for the bosonic  state in Eq.~(\ref{CS6}),   identical   arguments apply also  to the fermionic state in Eq.~(\ref{fermionCS6}).

\section{Derivations of the  results  of Section \ref{sec5} and  proof of  Theorem 4  and Proposition 3}
\label{appD}
Due to the invariance of $\hat\Pi_{\bm m}$ under permutations of the particle slots   we can  consider the    occupation vector $\bm{m}^\prime$, $|\bm{m}^{\prime}|=r$,  to be represented by the first $r$ entries $l_1,\ldots,l_r$ of an output-mode sequence. Then the  summation over $\bm{m}^{\prime\prime}:=\bm{m}-\bm{m}^\prime $, $|\bm{m}^{\prime\prime} | = n-r$   can be performed by converting it into a summation over $l_{r+1},\ldots, l_n$ using Eq.~(\ref{sum_ID}) of Section \ref{sec5}. From Eq.~(\ref{DetOp}) of Section \ref{sec2} we get 
\begin{eqnarray}
\label{D1}
&&\hat{\Pi}^{(r|n)}_{\bm{m}^\prime} =   \binom{n}{r}^{-1} \sum_{\bm{m}^{\prime\prime} } \frac{\bm{m}!}{\bm{m}^\prime! \bm{m}^{\prime\prime} !}\hat{\Pi}_{\bm{m}^\prime+\bm{m}^{\prime\prime} }  
= \frac{r!}{n!\bm{m}^\prime !} \sum_{l_{r+1}=1}^M \ldots \sum_{l_n=1}^M \sum_{\sigma\in S_n} \hat{P}_\sigma \bigotimes_{\alpha=1}^n |l_\alpha\rangle\langle l_\alpha| \hat{P}^\dag_\sigma =  \nonumber\\
&& = \frac{r!}{n! \bm{m}^\prime!} \sum_{\sigma\in S_n} \hat{P}_\sigma \left[ \bigotimes_{\alpha=1}^r |l_\alpha\rangle\langle l_\alpha | \otimes \hat{I}^{\otimes n-r}\right] \hat{P}^\dag_\sigma = \frac{r!(n-r)!}{n! \bm{m}^\prime!} \sum_{\nu\in S_{r|n}} \hat{P}_\nu \left[ \sum_{\tau\in S_r} \hat{P}_\tau\left(\bigotimes_{\alpha=1}^r |l_\alpha\rangle\langle l_\alpha | \right)\hat{P}^\dag_\tau \otimes \hat{I}^{\otimes n-r}\right] \hat{P}^\dag_\nu  \nonumber\\
&&=\binom{n}{r}^{-1} \sum_{\nu\in S_{r|n}} \hat{P}_\nu\left[\hat{\Pi}_{\bm{m}^\prime} \otimes \hat{I}^{\otimes n-r}\right] \hat{P}^\dag_\nu
\end{eqnarray}
where   we have used the  decomposition of $\sigma$  as follows
\[
\sigma = \nu(\tau\otimes \pi),\quad \tau \in S_r,\quad \pi \in S_{n-r},\quad \nu \in S_{r|n}:=\frac{S_n}{S_r\times S_{n-r}},
\]
  (the summation over $\pi$  gives the factor $(n-r)!$) and we have  introduced the    $r$-particle detection operator  
 \be
\hat{\Pi}_{\bm{m}^\prime}=  \frac{1}{ \bm{m}^\prime!}\sum_{\tau \in S_r} \hat{P}_{\tau}\left( \bigotimes_{\alpha=1}^r |\tilde{l}_\alpha\rangle\langle \tilde{l}_\alpha | \right) \hat{P}^\dag_{\tau}. 
 \en{D2}
The permutation $\nu$ in Eq.~(\ref{D1})  selects $r$ out of $n$  single-particle Hilbert spaces to be acted on by the $r$-particle detection operator in Eq.~(\ref{D2}).

Now let us show that the   marginal probability  of detecting $r$ out of $n$ particles in the occupation vector $\bm{m}^\prime$ over   $b$-modes Eqs.~(\ref{Uab2})-(\ref{Uab1})  of Section \ref{sec3} reads
 \be
 p^{(r|n) }_{\bm{m}^{\prime},\bm{n}} = \mathrm{Tr}\left( \hat{\Pi}^{(r|n)}_{\bm{m}^\prime} \hat{\varrho}_{(vis)}\right) =  \mathrm{Tr}\left( \hat{\Pi}_{\bm{m}^\prime} \hat{\varrho}^{(r|n)}_{(vis)}\right)
 \en{D3}
 with the marginal $r$-particle  visible state  
\be
  \hat{\varrho}^{(r|n)}_{(vis)} =  \mathrm{Tr}_{{r+1},\ldots,n}{\hat{\varrho}_{(vis)}},
 \en{D4}
where, by the particle-permutation symmetry of the visible state,  we can  trace over an  arbitrary subset  of  $n-r$ single-particle Hilbert spaces in $\mathcal{H}^{\otimes n}_{(vis)}$.  To show Eqs.~(\ref{D3})-(\ref{D4})  let us  use the following  decompositions of permutations:
\[
\sigma = \nu({\sigma}_1\otimes {\sigma}_2),\quad \pi = \mu({\pi}_1\otimes {\pi}_2), \quad \nu(1,\ldots, n) = (\alpha_1,\ldots,\alpha_n), \quad \mu(1,\ldots,n) = (\beta_1,\ldots, \beta_n),
\]
where we have introduced order-preserving  $(r,n-r)$-shuffles $\nu,\mu\in \frac{S_n}{S_r\times S_{n-r}}$, whereas  ${\sigma}_j,{\pi}_j$ act on the first $r$ ($j=1$) or last $n-r$ ($j=2$) elements.   The summation over shuffles $\nu$ and $\mu$ is therefore equivalent to summation over all choices of the first $r$ indices  $\bm{\alpha} = (\alpha_1,\ldots,\alpha_r)$ and $\bm{\beta}=(\beta_1,\ldots,\beta_r)$.

With these definitions, we can   rewrite the visible state  of $n$ identical particles in Eq.~(\ref{vis_state}) of Section \ref{sec3} as follows
\begin{eqnarray}
\label{D6}
 &&\hat{\varrho}_{(vis)}    = \frac{1}{n!\bm{n}!}\sum_{\bm{\alpha}} \sum_{\bm{\beta}}\sum_{\sigma_1,\pi_1\in S_r}\sum_{\sigma_2,\pi_2\in S_{n-r}}\Lambda (\pi\sigma^{-1}) \left[ \bigotimes\limits_{i=1}^r | k_{\alpha_{\sigma_1(i)}}\rangle\langle k_{\beta_{\pi_1(i)}}|\right]\otimes \left[ \bigotimes\limits_{i=r+1}^n | k_{\alpha_{\sigma_2(i)}}\rangle\langle k_{\beta_{\pi_2(i)}}|\right].\nonumber\\
 \end{eqnarray} 
Since the  visible state is particle-permutation symmetric, we can omit the averaging over $\nu\in S_{r|n}$ in the  operator $\hat{\Pi}^{(r|n)}_{\bm{m}^\prime}$ in Eq.~(\ref{D1}) when using it in Eq.~(\ref{D3}). Therefore,  we can first take the trace over the last $n-r$ single-particle Hilbert spaces in the expression for the visible state in Eq.~(\ref{D6}). Taking the latter trace gives us the  marginal visible state defined in Eq.~(\ref{D4}). To  find a closed expression  for  it,  we observe the  following identity 
\be
 \sum_{\bm{\beta}} \prod_{i=r+1}^n \langle k_{\beta_{\pi_2(i)}}|k_{\alpha_{\sigma_2(i)}}\rangle = \frac{\bm{n}!}{\bm{n}^\prime! \bm{n}^{\prime\prime}!}\sum_{\tau_2\in \mathcal{Y}_{\bm{n}^{\prime\prime}}}\delta_{\pi_2\sigma_2^{-1},\tau_2 },
\en{D7}
where $\bm{n}^\prime$ and $\bm{n}^{\prime\prime}$  are occupation vectors corresponding to the sequences of modes $k_{\alpha_1},\ldots,k_{\alpha_r}$ and $k_{\alpha_{r+1}},\ldots,k_{\alpha_n}$ and $\mathcal{Y}_{\bm{n}^{\prime\prime}} =S_{{n}^{\prime\prime}_1}\times \ldots \times S_{{n}^{\prime\prime}_M}$ is the Young subgroup for the occupation vector  $\bm{n}^{\prime\prime}$. The  inner product in Eq.~(\ref{D7})    is   non-zero  only when   the shuffled multisets of input modes coincide 
\be
\{k_{\beta_1},\ldots, k_{\beta_r}\} = \{k_{\alpha_1},\ldots, k_{\alpha_r}\},
\en{k_set}
i.e.,    $\tau:=\mu\nu^{-1}   \in \mathcal{Y}_{\bm{n}}$,  and    $\tau_2:=\pi_2\sigma_2^{-1}\in \mathcal{Y}_{\bm{n}^{\prime\prime}}$. We can therefore rewrite the relative permutation in   Eq.~(\ref{D6}) as follows
\be
\pi\sigma^{-1} = \tau  \nu\left( {\pi}_1{\sigma}_1^{-1}\otimes{\pi}_2{\sigma}_2^{-1}\right) \nu^{-1}. 
\en{rel_per} 
  We can use  only one of the   coinciding multisets of Eq.~(\ref{k_set}) in both the ket- and bra-states in Eq.~(\ref{D6}) without affecting the result.  Due to    the symmetry of the label state  $\hat{\varrho}^{(l)}$ with respect to the Young subgroup (see Eq.~(\ref{ED13}) of Appendix  \ref{appB}),  the   indistinguishability function in this case simplifies accordingly:
\begin{eqnarray}
\label{Lambda_nprime}
\Lambda(\pi\sigma^{-1}) &=&  \vare(\pi\sigma^{-1}) \mathrm{Tr}\left(\hat{P}_{\pi\sigma^{-1}} \hat{\varrho}^{(l)}\right) =\vare(\pi_1\sigma_1^{-1})  \vare(\tau_2)\mathrm{Tr}\left(\hat{P}_\nu\left[\hat{P}_{\pi_1\sigma^{-1}_1}\otimes \hat{P}_{\tau_2}\right]  \hat{P}_{\nu^{-1}}\hat{\varrho}^{(l)}\right)
\nonumber\\
 &\equiv& \vare(\pi_1\sigma_1) J_{\bm{n}^\prime}(\pi_1\sigma_1^{-1})
\end{eqnarray}
where  we have introduced  the  indistinguishability function $J_{\bm{n}^\prime}$ on $S_r$ for  the subset of  $r$ identical particles in the occupation vector $\bm{n}^\prime$  corresponding to    $k_{\alpha_1},\ldots, k_{\alpha_r}$  shuffled by   $\nu$.   With these observations we get  the following expression 
 \begin{eqnarray}
\label{D8}
 &&\hat{\varrho}^{(r|n)}_{(vis)}    =  \binom{n}{r}^{-1} \sum_{\bm{\alpha}} \frac{1}{r!\bm{n}^\prime!}\sum_{\sigma_1,\pi_1\in S_r} \vare(\pi_1\sigma_1) J_{\bm{n}^\prime}(\pi_1\sigma_1^{-1})  \bigotimes\limits_{i=1}^r | k_{\alpha_{\sigma_1(i)}}\rangle\langle k_{\alpha_{\pi_1(i)}}|,
 \end{eqnarray} 
We have arrived at the result in Eq.~(\ref{D3}) and have found the form of the marginal state there. Comparing with the visible state in Eq.~(\ref{vis_state}) of Section \ref{sec3} we obtain the marginal probability given  by Eq.~(\ref{marprob2}) of Section \ref{sec5}. 

\subsection{Proof of Theorem 4}
 For $1\le r\le n$, let $\hat S^{(\pm)}$ act on the first $r$  factors  in the visible Hilbert space  $\mathcal{H}_{(vis)}^{\otimes n}$   and define
\[
 \hat{Q}_r=\hat S^{(\pm)}\otimes\hat I^{\otimes(n-r)}.
\]
By the definition of the partial trace,
$\mathcal D^{(r|n)}=\mathrm{Tr}(\hat{Q}_r\hat\varrho_{(vis)})$.
For $r_1<r_2$, the range of $\hat{Q}_{r_2}$ lies in the range of $\hat{Q}_{r_1}$, for either choice of statistics. Consequently,
\be
 \hat{Q}_{r_2}\hat{Q}_{r_1}=\hat{Q}_{r_1}\hat{Q}_{r_2}=\hat{Q}_{r_2},
 \qquad \hat{Q}_{r_2}\le \hat{Q}_{r_1}.
 \en{D9}
Taking the trace against the positive semidefinite state gives
\be
 \mathcal D^{(r_2|n)}
 =\mathrm{Tr}(\hat{Q}_{r_2}\hat\varrho_{(vis)})
 \le\mathrm{Tr}(\hat{Q}_{r_1}\hat\varrho_{(vis)})
 =\mathcal D^{(r_1|n)}.
 \en{D10}
This proves the hierarchy directly for the reduced visible states; permutation invariance makes the particular choice  of    single-particle Hilbert spaces  in  the $n$th tensor power of the visible Hilbert space  $\mathcal{H}_{(vis)}^{\otimes n}$ immaterial. Q.E.D.

\subsection{Proof of Proposition 3}

Consider first an $n$-particle label state given by a single tensor power $\hat{\varrho}^{(l)}_n=\hat{\rho}^{\otimes n}$. 
 Let $\{q_i\}$ denote the set of eigenvalues of $\hat\rho$. Define 
\be
p_k = \mathrm{Tr}\hat{\rho}^k = \sum_i q^k_i.
\en{pkDEF}
  Using Eq.~(\ref{DJ}) of Section \ref{sec3} and Eq.~(\ref{J_tensor}) of Section \ref{sec4} we obtain the   following  sum for the projective measure,
 \be 
      \mathcal{D}_n|_{\hat{\rho}^{\otimes n}}   =    \frac{1}{n!}\sum_{\sigma\in S_n} \prod_{k=1}^n p_k^{C_k(\sigma)}. 
\en{Dncycle}
To compute the cycle sum we can use the standard generating function for the cycle sum  \cite{Stanley} 
\be
\Phi(t):=\sum_{n=0}^\infty \mathcal D_n   t^n = \exp\left\{ \sum_{k=1}^\infty \frac{p_kt^k}{k}\right\} = \exp\left\{\sum_i  \sum_{k=1}^\infty \frac{q_i^kt^k}{k}\right\} 
= \prod_i \frac{1}{1-t q_i},
\en{genFUN}
where we set $\mathcal D_0:=1$. The identities hold for $|t|<1/\|\hat\rho\|$ (or coefficientwise as formal power series); the nonnegative trace-class spectrum makes the coefficient sums well defined. 
By extracting the     coefficient of $t^n$ on both sides of Eq.~(\ref{genFUN}),  we obtain 
\be
 \mathcal D_n   = \sum_{i_1\le \ldots \le i_n} q_{i_1}\ldots q_{i_n}, 
 \en{Dsympoly}
i.e.,  the complete homogeneous symmetric polynomial of degree $n$. By taking the logarithmic derivative of both sides of Eq.~(\ref{genFUN}) we get
\be
\Phi^\prime(t) = \Phi(t) \sum_{k=1}^{\infty}p_k t^{k-1}. 
\en{logder}
Comparing the  coefficients of $t^{n-1}$ on both sides of Eq.~(\ref{logder}) we obtain the   recurrence identity
\be
n \mathcal D_n = \sum_{k=1}^n p_k \mathcal D_{n-k} .
\en{recurr}
Introduce a random variable $X$,  which takes the value $q_i$ with probability $q_i$. Then,   Jensen's inequality applied to  $p_k = \langle X^{k-1}\rangle$ (see Eq.~(\ref{pkDEF}))  gives 
\be
p_k=  \langle X^{k-1}\rangle  \ge  \langle X\rangle^{k-1} =  p_2^{k-1}, \quad \forall k\ge 2.
\en{pkbyp2}
The bound in Eq.~(\ref{pkbyp2}) also holds for $k=1$, since $p_1=1$. Using Eqs.~(\ref{recurr})-(\ref{pkbyp2}), we can prove by induction the result in Proposition 3 of Section \ref{sec5}. First, by  Eq.~(\ref{Dncycle})  
\[
\mathcal{D}_2 = \frac{1+ p_2}{2}.
\]
Thus we need to prove that
\be
\mathcal D_n \ge p_2^{n-1}.
\en{Dnbyp2}
For $n=1$ we have $\mathcal D_1 = 1$ and $p_1 = 1$. Assume that $\mathcal D_m \ge p_2^{m-1}$ for $1\le m <n$. Then, by rewriting Eq.~(\ref{recurr}) and using the assumption and the fact that $p_2\le 1$,     we get
\[
n\mathcal D_n = p_n + \sum_{k=1}^{n-1}p_k \mathcal D_{n-k}\ge p_2^{n-1} + \sum_{k=1}^{n-1}p_2^{k-1}p_2^{n-k-1} \ge  p_2^{n-1} +(n-1) p_2^{n-2} \ge  n p_2^{n-1}.
\]
Thus, the proposition is proven for an $n$-particle label state given by the $n$th tensor power of the single-particle label state. 

Finally, for an $n$-particle label state given by a convex mixture,  by using the above result for a tensor-power label state and Jensen's inequality  we get
\[
\mathcal D_n = \int \mathcal{D}_n|_{\hat{\rho}^{\otimes n}}  d\mu (\hat{\rho})\ge  \int(\mathrm{Tr}\hat{\rho}^{2})^{n-1} d\mu (\hat{\rho})\ge  \left(\int\mathrm{Tr}\hat{\rho}^{2} d\mu (\hat{\rho})\right)^{n-1} = \left( 2\mathcal{D}_2-1\right)^{n-1}.
\]
 Moreover, by taking the partial  trace we have $\mathcal{D}^{(2|n)} = \mathcal{D}_2$. Q.E.D.

\twocolumngrid


\begin{thebibliography}{9}
\bibitem{HOM} C. K. Hong, Z. Y. Ou, and L. Mandel, 
Measurement of subpicosecond time intervals between two photons by interference, 
Phys. Rev. Lett. \textbf{59}, 2044 (1987). 


\bibitem{ElecHOM} R. C. Liu, B. Odom, Y. Yamamoto, and  S. Tarucha,
Quantum interference in electron collision,
Nature \textbf{391}, 263 (1998). 


\bibitem{AtomHOM} R. Lopes, A. Imanaliev, A. Aspect, M. Cheneau,  D. Boiron and C. I. Westbrook,  
Atomic Hong–Ou–Mandel experiment, 
Nature \textbf{520}, 66 (2015). 

\bibitem{TwFerQW} L. Sansoni, F. Sciarrino, G.  Vallone, P. Mataloni, A. Crespi, R. Ramponi, and R. Osellame, 
Two-Particle Bosonic-Fermionic Quantum Walk via Integrated Photonics, 
Phys. Rev. Lett. \textbf{108}, 010502 (2012). 

\bibitem{Ou1} Z. Y. Ou, 
Temporal distinguishability of an $N$-photon state and its characterization by quantum interference, 
Phys. Rev. A \textbf{74}, 063808 (2006). 
\bibitem{Ou2} Z. Y. Ou, 
Characterizing temporal distinguishability of an $N$-photon state by a generalized photon bunching effect with multiphoton interference, 
Phys. Rev. A \textbf{77}, 043829 (2008). 

\bibitem{GenHOM} Y. L.  Lim and A.  Beige, 
Generalized Hong–Ou–Mandel experiments with bosons and fermions, 
New J. Phys. \textbf{7},  155 (2005). 

\bibitem{SymBeyBS} M. C. Tichy, M. Tiersch, F. Mintert, and A. Buchleitner, 
Many-particle interference beyond many-boson and many-fermion statistics, 
New J. Phys. \textbf{14}, 093015 (2012).

\bibitem{ZeroTranSymm} C.  Dittel,  G.  Dufour,  M.  Walschaers, G.  Weihs,  A. Buchleitner,  and R. Keil, 
Totally destructive interference for permutation-symmetric many-particle states, 
Phys. Rev. A \textbf{97}, 062116 (2018).

\bibitem{Symm4Dist} J.  M\"unzberg,  C.  Dittel, M. Lebugle,  A.  Buchleitner, A. Szameit,  G. Weihs,  and R. Keil, 
Symmetry Allows for Distinguishability in Totally Destructive Many-Particle Interference, 
PRX Quantum 2, 020326 (2021).


\bibitem{MetHOM} E. Descamps,  A. Keller,  and P. Milman, 
The Role of Symmetry in Generalized Hong-Ou-Mandel Interference and Quantum Metrology, Phys. Rev. Lett. \textbf{136}, 060807 (2026). 

 

 \bibitem{LF1} R. Lo Franco and G. Compagno, Quantum entanglement of identical particles by standard information-theoretic notions, Sci. Rep.  \textbf{6}, 20603 (2016).


\bibitem{LF2} G. Compagno, A. Castellini, and R. Lo Franco, Dealing with indistinguishable particles and their entanglement, Phil. Trans.  R. Soc.  A \textbf{376}, 20170317 (2018).
 
\bibitem{LF3}  R. Lo Franco and G. Compagno, Indistinguishability of elementary systems as a resource for quantum information processing, Phys. Rev.  Lett. \textbf{120}, 240403 (2018).

\bibitem{RepEnt} F. Benatti, R. Floreanini, F. Franchini and U. Marzolino, Entanglement in indistinguishable particle systems, Phys. Rep. \textbf{878}, 1 (2020). 

\bibitem{Shch2015} V. S. Shchesnovich, 
Partial indistinguishability theory for multiphoton experiments in multiport devices, 
Phys. Rev. A \textbf{91}, 013844 (2015).


\bibitem{Tch2015} M. C. Tichy, 
Sampling of partially distinguishable bosons and the relation to the multidimensional permanent,
Phys. Rev. A \textbf{91}, 022316 (2015).


\bibitem{WeylD} M. Tillmann, S.-H. Tan, S. E. Stoeckl, B. C. Sanders, H. de Guise, R. Heilmann, S. Nolte, A. Szameit, and  P.~Walther, 
Generalized Multiphoton Quantum Interference,
Phys. Rev. X \textbf{5}, 041015 (2015).

\bibitem{NonMon4ph}  Y.-S. Ra, M. C. Tichy, H.-T. Lim, O. Kwon, F. Mintert, A. Buchleitner, Y.-H. Kim, 
Nonmonotonic quantum-to-classical transition in multiparticle interference. 
Proc. Natl. Acad. Sci. U.S.A. \textbf{110}, 1227  (2013).

\bibitem{3phPhase} A. J. Menssen, A. E. Jones, B. J. Metcalf, M. C. Tichy, S. Barz, W. S. Kolthammer, and I. A. Walmsley, 
Distinguishability and Many-Particle Interference, 
Phys. Rev. Lett. \textbf{118}, 153603 (2017). 

\bibitem{DistMix3ph} A. E. Jones,  S. Kumar,  S. D’Aurelio, M. Bayerbach, A. J. Menssen,  and S.  Barz, 
Distinguishability and mixedness in quantum interference, 
Phys. Rev. A  \textbf{108}, 053701 (2023). 

\bibitem{nphPhases} V. S. Shchesnovich and M. E. O. Bezerra, 
Collective phases of identical particles interfering on linear multiports,
Phys. Rev. A \textbf{98}, 033805 (2018).

\bibitem{DistPhInter} A. E. Jones, A. J. Menssen, H. M. Chrzanowski, T. A. W. Wolterink, V. S. Shchesnovich,
I. A. Walmsley, 
Multiparticle interference of pairwise distinguishable photons. 
Phys. Rev. Lett. \textbf{125}, 123603 (2020).


\bibitem{MultPhInd} M. Pont, R. Albiero, S. E. Thomas, N. Spagnolo, F. Ceccarelli, G. Corrielli, \textit{et al.},
 Quantifying $n$- Photon Indistinguishability with a Cyclic Integrated Interferometer. 
Phys. Rev. X \textbf{12}, 031033 (2022).

\bibitem{Distchar} S. N. van den Hoven, M. C. Anguita, S. Marzban, and J.~J. Renema, 
Quantum Advantage for Single-Photon State Characterization,
Phys. Rev. Lett. \textbf{137}, 073604  (2026). 

 \bibitem{VS2016} V. S. Shchesnovich, 
Universality of Generalized Bunching and Efficient Assessment of Boson Sampling, 
Phys. Rev. Lett. \textbf{116}, 123601 (2016). 


\bibitem{BCount}  B. Seron, L. Novo and N. J. Cerf, Boson bunching is not maximized by indistinguishable particles, Nat. Photon. \textbf{17}, 702 (2023).


\bibitem{Geller2026} S.  Geller  and E.  Knill, 
Measuring multiparticle indistinguishability with the generalized bunching probability, 
Phys. Rev. A \textbf{113}, 042606 (2026).

\bibitem{Shch2014} V. S. Shchesnovich, 
Sufficient condition for the mode mismatch of single photons for scalability of the boson-sampling computer, 
Phys. Rev. A \textbf{89}, 022333 (2014).

\bibitem{DistNew} M. Englbrecht,  T.  Kraft, C.  Dittel,   A.  Buchleitner, G.  Giedke  and B.  Kraus, 
Indistinguishability of Identical Bosons from a Quantum Information Theory Perspective,
Phys. Rev. Lett.  \textbf{132}, 050201 (2024). 


\bibitem{Shch2015A} V. S. Shchesnovich, 
Tight bound on the trace distance between a realistic device with partially indistinguishable bosons and the ideal BosonSampling,  
Phys. Rev. A \textbf{91},   063842  (2015). 

\bibitem{AA} S. Aaronson and A. Arkhipov, 
The computational complexity of linear optics, 
Theory of  Computing \textbf{9},  143 (2013).

\bibitem{20ph60mod} H.  Wang, J.  Qin, X. Ding, M.-C. Chen, S. Chen, X. You  \textit{et al.},
Boson Sampling with 20 Input Photons and a 60-Mode Interferometer in a $10^{14}$-Dimensional Hilbert Space,   
Phys. Rev. Lett. \textbf{123},  250503 (2019).
 
\bibitem{SimBSdist} J. J. Renema, A. Menssen, W. R. Clements, G. Triginer, W. S. Kolthammer, and I. A. Walmsley, 
Efficient Classical Algorithm for Boson Sampling with Partially Distinguishable Photons. 
Phys. Rev. Lett. \textbf{120}, 220502 (2018).


\bibitem{LOC} E. Knill, R. Laflamme and G. J. Milburn, 
A scheme for efficient quantum computation with linear optics, 
Nature \textbf{409}, 46 (2001). 

\bibitem{RevLOC} P. Kok, W. J. Munro, K. Nemoto, T. C. Ralph, J. P. Dowling and G. J. Milburn, 
Linear optical quantum computing with photonic qubits, 
Rev. Mod. Phys. \textbf{79}, 135 (2007). 

\bibitem{LecNotes} V.S. Shchesnovich, \textit{The second quantization method for indistinguishable particles}, arXiv:1308.3275.	
	
	



 \bibitem{BookNC} M. A. Nielsen and  I. L. Chuang, \textit{Quantum Computation and Quantum Information}, Quantum Information: 10th Anniversary Edition, 10th ed.
(Cambridge University Press, New York, NY, 2011). 
 
 \bibitem{BSF} V. S. Shchesnovich, 
 Boson-sampling with non-interacting fermions, 
 Int.  Journal of Quant.  Inf. \textbf{13},   1550013 (2015). 


\bibitem{PartDistInv} E. Annoni  and S. C. Wein, 
Incoherent behavior of partially distinguishable photons, 	
arXiv:2502.05047. 



\bibitem{GB2026} G. Dufour and A. Buchleitner, 
Fourier analysis of many-body transition amplitudes and states, 
 New J. Phys. \textbf{28},  084515 (2026). 



\bibitem{MBCoh22} E. Brunner,  A. Buchleitner, and G. Dufour, 
Many-body coherence and entanglement probed by randomized correlation measurements,
Phys. Rev. Research \textbf{4}, 043101 (2022). 


\bibitem{Stanley} R. P. Stanley, \textit{Enumerative Combinatorics}, 2nd ed., Vol. 1 (Cambridge University Press, 2011).



\end{thebibliography}
\end{document}